\documentstyle[12pt,vi]{mitthesis}
\begin{document}

\def\half{\frac{1}{2}}
\input BoxedEPS
\SetRokickiEPSFSpecial  
\HideDisplacementBoxes

%
%
%
%
\title{Exact Renormalized One-Loop Quantum Corrections to Energies of
Solitonic Field Configurations}

\author{Noah Matthew Graham}
\department{Department of Physics}
\degree{Doctor of Philosophy}
\degreemonth{June}
\degreeyear{1999}
\thesisdate{April 30, 1999}


\supervisor{Edward Farhi}{Professor}

\chairman{Thomas J. Greytak}{Associate Department Head for Education}

\maketitle



\newpage
\setcounter{savepage}{\thepage}
\begin{abstractpage}
%
%
%
We develop a method for computing exact one-loop quantum corrections to the
energies of static classical backgrounds in renormalizable quantum
field theories.  We use a continuum density of states formalism to
construct a regularized Casimir energy in terms of phase shifts and
their Born approximations.  This method unambiguously incorporates
definite counterterms fixed in the standard way by physical
renormalization conditions.  The result is a robust computation that can
be efficiently implemented both numerically and analytically.  We
carry out such calculations in models of bosons and fermions in one
and three dimensions.

\end{abstractpage}


\newpage

\section*{Acknowledgments}

I would like to thank Professors Edward Farhi and Robert L. Jaffe for
being outstanding teachers and collaborators, providing constant
support, guidance and encouragement throughout this project.  I would
also like to thank Benjamin Scarlet for computational help, especially for
suggesting a crucial speedup to the calculation of higher order Born
approximations.  I also would like to thank Sergei Bashinsky, Oliver
DeWolfe, Kieran Holland, Tamas Hauer, and Herbert Weigel for many valuable
discussions.  And my deepest thanks to Nancy, Daniel, and my parents.

\pagestyle{plain}
\tableofcontents
\newpage
\listoffigures

\chapter{Introduction}

Given a static field configuration in a renormalizable quantum field
theory, a natural question to ask is what its energy is.  This
question is especially relevant in the study of solitons, since they are
local minima of the energy.  Classically, the energy is simple to
calculate from the Lagrangian density of the theory.  However,
quantization of the theory introduces corrections to this classical
result.  Such corrections can be expanded as a power series in the
coupling constants of the theory, which is equivalent to an expansion
in $\hbar$.

In this work we will consider the leading quantum corrections to the
classical energies of field configurations in a variety of quantum
field theories.  Since the power of $\hbar$ counts the number of loops in
the diagrammatic expansion of the energy, taking the leading correction
will correspond to a one-loop calculation, in which we sum all
one-loop diagrams with arbitrary numbers of insertions of the
classical background field.  This calculation is equivalent to summing
the shifts in the $\half \hbar \omega $ zero-point energies of all the
small fluctuations modes in the presence of the background field.
This sum is known as the Casimir energy.  As usual in quantum field
theory, the result of such a calculation diverges, and we must
introduce divergent counterterms that are fixed through a finite set
of renormalization conditions.  The renormalization conditions define
the theory in terms of physical quantities.  Our challenge will be to implement
such a calculation in a robust, efficient and unambiguous way.  In
particular, we must isolate the cancellation of divergent quantities
without missing any finite contributions, and we must be sure that we
have implemented the renormalization conditions faithfully.  Merely
``canceling the infinities'' is clearly not sufficient when we want to
regard a finite result as a physical prediction of a particular theory
defined under fixed renormalization conditions.

One application of this method is in coupling of the Higgs sector of
the Standard Model to heavy quarks.  If we imagine adjusting the
Yukawa coupling for a quark doublet so that the quarks' mass becomes
very large, we would expect that when they become sufficiently heavy,
they effectively decouple from the theory.  However, this cannot be
the whole story, since if we simply removed a quark doublet from the
Standard Model, we would ruin anomaly cancellation.  The resolution of
this paradox \cite{Farhi} requires that solitons in the Higgs field
carry the quantum numbers of the decoupled fermions.  This result
suggests a picture in which heavy quarks are realized at small coupling
as elementary fermion excitations, and at large coupling as Higgs
solitons.  In between these two limits, one might then expect to find
a hybrid configuration, with the heavy quark tightly bound to a
deformation in the Higgs field.  To see if this picture is correct, we
need to do a variational computation:  we must find the field
configuration of lowest energy that carries the heavy quark quantum numbers. 
This application highlights the importance of unambiguously fixing our
renormalization conditions and avoiding finite errors in our energy
computation.  If our renormalization conditions were not fixed
precisely, we would effectively be changing the theory as we moved from
one background field to another, rendering the variational calculation
meaningless.

One could imagine building a Higgs configuration that brings a heavy
quark level down from the mass of the quark closer to (or below) zero.
The Higgs configuration itself would have an energy cost from the gradient
and potential terms in the classical bosonic Hamiltonian.  This cost could be
balanced by the shift in the ``valence'' quark energy level, since the
quark occupying this state moves to a lower energy.  In particular,
for strong quark Yukawa coupling (which means large quark mass), and
small Higgs self-coupling (which means small Higgs mass), deforming
the Higgs field would be favored.  However, these two pieces alone do
not form a self-consistent semiclassical calculation.  We have
included the classical energy along with one part of the leading
quantum correction to that energy, the shift in the valence level.
We have no justification for ignoring the shifts in all the
other quark levels, which contribute at the same order through their
zero-point energies.  Thus we are forced consider the full Casimir
energy in this problem.  If we do include the Casimir energy, we
obtain a self-consistent semiclassical result.  One may doubt the
validity of the semiclassical approximation as the coupling gets
large, of course, but we will obtain a result that is valid in a
well-defined approximation.  (In some models, it is also the exact
result in a large-$N$ limit \cite{BN}.) 

Although we must compute the full Casimir energy, not just the valence
contribution, it is still possible to construct self-consistent approximations
to this quantity.  Of course, these approximations further restrict
the domain of validity of the computation.  In particular, if the
background field is slowly varying on the scale of the Compton
wavelength of the quantum fluctuations, the derivative expansion
becomes valid.  Indeed, it has been used in models similar to the ones
we will consider \cite{BN}.  However, in both models of heavy quarks
and other models we will want to explore, the scale at which the
background field varies is precisely the Compton wavelength of the
quantum fluctuations, so that all the terms in the derivative
expansion will be about the same size, rendering the expansion unreliable.
Our approach will be exact to one-loop order.

Central to our technique will be to re-express the Casimir sum in
the continuum in terms of phase shifts, using a formalism that
originates with Schwinger's work on QED in the presence of strong
fields \cite{Schw}.  We will consider only field
configurations with some form of spherical symmetry, so that we can
use a partial-wave decomposition.  In each channel, the difference
between the free and interacting density of states is related to the phase 
shift by
\begin{equation}
 \rho(k) = \rho_0(k) + \frac{1}{\pi} \frac{d\delta(k)}{dk}
\end{equation}
which follows from imposing a boundary on the system and then sending
the boundary to infinity, or by more formal S-matrix arguments.
In any channel, for any $k$, the phase shift is a finite,
well-defined, physical quantity.  Its analytic structure is well
understood in terms of Jost functions, and it can also be rigorously
related to the Green's functions and S-matrix of the theory.  The sum
over continuum modes is then replaced by an integral over the density
of states
\begin{equation}
E = \half\sum_j \omega_j + \half \int_0^\infty dk \sqrt{k^2 + m^2} \rho(k)
\end{equation} 
in which the bound states are still included explicitly.  This
mathematical artillery leads to several important properties of the
phase shifts: 
\begin{itemize}
\item
We can expand the phase shifts as a Born series in the strength of the
potential.  This expansion is in exact correspondence with the
expansion of the full propagator of the theory in terms of the free
propagators connecting insertions of the potential.  The Born
expansion also has simple behavior at large $k$ and becomes more and
more accurate in this limit, which will enable us to use it as a
regulator of ultraviolet divergences.
\item
The phase shifts track level crossings, ensuring a consistent counting
of modes as states become bound.  This property allow us to avoid a serious
problem we would encounter if we put the system in a box, because in
that case we would have a hard time ensuring that we have kept the
appropriate number of modes when computing the Casimir energy and the
contribution of the counterterms.  Missing even one mode in the
Casimir sum will lead to a drastic change in the final answer;
although such an error is small compared to the leading  (divergent)
behavior of the sum, the leading behavior is cancelled by the
counterterms.  The final answer is generically of the same order as
a typical energy level.  Our key tool here will be Levinson's theorem,
which relates the phase shift at $k = 0$ to the number of bound states. 
\item
The phase shifts and their Born approximations are simple, robust
quantities amenable to both numerical and analytic calculation.  We
will see that they enable us to replace the cancellations of large
quantities, the Casimir energy against the counterterms, by the much more
manageable cancellation of exact phase shift against Born
approximation.  Many cases will in the end be tractable only
numerically, but we will also find that numerical analysis will also
shed light on analytic results by allowing us to continuously
interpret between a trivial configuration and one that can be solved
analytically.
\item
The absence of boundaries will allow us to avoid spurious
contributions from artificial boundary conditions.  This property will
be especially useful in models with fermions.
\item
In principle, phase shifts can even be computed in fractional
dimensions, since they obey a simple radial differential equation that
depends analytically on the dimension of space.  In particular, the
first Born approximation the bosonic phase shifts can be analytically
calculated in arbitrary dimensions terms of generalized Bessel
functions.  This result agrees exactly with the result one would find for the
corresponding tadpole graph evaluated in dimensional regularization.

\end{itemize}

Much of this work originally appeared in \cite{us}.

\clearpage
\newpage

\chapter{$\phi^4$ theory in 1+1~dimensions}

We will begin with examples in 1+1~dimensions.  The study of exactly
soluble 1+1 dimensional problems has yielded many insights into fundamental
problems in field theory.  Other 1+1 dimensional problems cannot be solved
exactly, making it important to understand which properties of exact
results will generalize to more generic cases and which are special to the
exactly soluble cases.  The renormalization process is simpler in one
dimension than it is in three, since theories in lower dimensions are less
divergent, but it still must be approached carefully.  Indeed, we will find
that the one dimensional problem actually contains additional subtleties
not present in higher dimensions.

\section{Formalism}

We will consider a standard $\phi^4$ theory
with spontaneous symmetry breaking and a source $J(x)$.  The action is
\begin{equation}
S[\phi] = \frac{m^2}{\lambda} \int
\left(\half(\partial_\mu \phi)^2 -  \frac{m^2}{8}(\phi^2 - 1)^2
  - J(x)\phi + {\cal L}_{\rm ct} \right) \, d^2x
\end{equation}
where ${\cal L}_{\rm ct}$ is the counterterm Lagrangian.  
Our metric has $g_{tt} = -g_{xx} = 1$.  We have rescaled the field $\phi$
somewhat unconventionally in order to make explicit the correspondence
between the powers of $\lambda$ and the powers of $\hbar$, which we
have set equal to 1.  Classical terms will go as $\frac{1}{\lambda}$,
the one-loop terms we compute will go as $\lambda^{0}$, and higher
loops will contribute with higher powers of $\lambda$.  The mass of
fluctuations around the trivial vacua $\phi(x)=\pm 1$ is $m$, and we
define the potential $U(\phi)=\frac{m^2}{8}(\phi^2 - 1)^2$.

We consider a fixed field configuration $\phi_0(x)$.  Initially we will
assume that $\phi_0(x) = \phi_0(-x)$, which restricts us to the
topologically trivial sector of the theory.  However, we will see that our
method works equally well for the case of $\phi_0(x) = -\phi_0(-x)$, so
that in fact we can deal with configurations with any topology as long as
$U(\phi_0)$ has reflection symmetry.  We adjust the source so that $\phi_0$ is
a stationary point of the action, which means that $J$ then solves the
equation
\begin{equation}
J(x) = \frac{d^2 \phi_0}{dx^2} - \half m^2 (\phi_0^3 - \phi_0).
\label{Jeq}
\end{equation}
We can always solve this equation for $J$, but there is no guarantee that the
$J$ we find will always correspond to a unique $\phi_0$, as we will see later.
If $\phi_0$ is a solution to the equations of motion, of course the source
will be zero.

We would like to consider the leading quantum correction to the classical
energy of this configuration, which we can represent as the sum of
the zero-point energies of the normal modes of small oscillations
around $\phi_0$.  Writing $\phi = \phi_0 + \eta$, the normal modes are
solutions of
\begin{equation}
-\frac{d^2 \eta}{dx^2} + (V(x)+m^2)\eta = \omega^2 \eta
\label{etaeq}
\end{equation}
with $V(x) = U''(\phi_0(x)) - m^2 = \frac{3}{2} m^2 (\phi_0^2(x) - 1)$.
The quantum change in energy in going from the trivial vacuum to
$\phi_0(x)$ is then
\begin{equation}
{\cal E}[\phi_0] = \half(\sum \omega - \sum \omega_0) + {\cal E}_{\rm ct}
= \Delta E + {\cal E}_{\rm ct}
\end{equation}
where $\omega_0$ are the free solutions (with $\phi_0^2(x) = 1$), and
${\cal E}_{\rm ct}$ is the contribution from the counterterms.

Note that it is possible that some of the values of $\omega^2$ will be
negative, so that in these directions our stationary point is a local
maximum rather than a minimum of the action.  These solutions will add an
imaginary part to the energy, which we can interpret via analytic
continuation as giving the decay rates through the unstable modes
\cite{Cole}. If there is a direction in field space in which small
oscillations lower the energy, we should be able to keep going in that
direction and arrive at a lower minimum.  Later, we will see explicitly
that the appearance of unstable modes coincides with the existence of a second
solution to the $\phi_0$ equation with the same $J$ and lower energy.

We would like to rewrite the sum over zero-point energies as an integral
over phase space of the product of the energy and the density of
states.  We can then break this integral into a sum over bound states
and an integral over a continuum, representing the latter in terms of
phase shifts.  In order to do so, however, we must review the
peculiarities of Levinson's theorem in one dimension.  For more
details on these results see \cite{Schiff} and \cite{lev}.  For a
symmetric $V(x)$, we can divide the continuum states into symmetric
and antisymmetric channels, and then calculate the phase shift as a
function of $k$ separately for each channel, where $\omega^2 = k^2 +
m^2$. The antisymmetric channel is completely equivalent to the $l =
0$ case in three dimensions, so we have
\begin{equation}
\delta_{\rm A}(0) = n_{\rm A} \pi
\end{equation}
where $n_{\rm A}$ is the number of antisymmetric bound
states.  However, we must be careful in dealing with the special case of a
state exactly at $k = 0$.  In this case the solution to eq.~(\ref{etaeq})
with $k = 0$ goes asymptotically to a constant as $x\to\infty$, as opposed
to the generic case where the $k=0$ solution goes to a constant plus linear
terms in $x$.  Just as in the $l = 0$ case in three dimensions, this state
contributes $\half$ to $n_{\rm A}$.  We will refer to such
states as ``half-bound states.''

In the symmetric channel, Levinson's theorem becomes
\begin{equation}
\delta_{\rm S}(0) = n_{\rm S} \pi - \frac{\pi}{2}
\end{equation}
where a bound state at $k = 0$ contributes $\half$ to $n_{\rm S}$, just as
in the antisymmetric case.  We can see the importance of getting the
half-bound states right by looking at the free case: the phase shift
is zero everywhere, and the  right-hand side is zero because the free case
has a half-bound state (the wavefunction $\psi = \rm{constant}$).  The
situation is equally subtle for reflectionless potentials, all of which
have half-bound states and $\delta_{\rm S}(0) + \delta_{\rm A}(0)$ equal to
an integer times $\pi$.

We are now ready to rewrite the change in the zero-point energies in terms
of phase shifts.  Letting $E_j$ be the bound state energies (again with $k = 0$
bound states contributing with a $\half$), $\rho(k)$ be the density of states
and $\rho_0(k)$ be the free density of states, we have
\begin{eqnarray}
\Delta E
&=& \half \sum_j E_j - \frac{m}{4} + \int_{0}^{\infty} \frac{dk}{2\pi}
\omega(k) (\rho(k) - \rho_0(k)) \nonumber \\
&=& \half \sum_j E_j - \frac{m}{4} + \int_{0}^{\infty} \frac{dk}{2\pi}
\omega(k) \frac{d}{dk}(\delta_{\rm A}(k) + \delta_{\rm S}(k))
\label{unrenorm}
\end{eqnarray}
where $\omega(k)=\sqrt{k^2+m^2}$ and we have used
\begin{equation}
\rho(k) = \rho_0(k) + \frac{1}{\pi} 
\frac{d}{dk}(\delta_{\rm A}(k) + \delta_{\rm S}(k)).
\end{equation}
Note that the $\frac{m}{4}$ term subtracts the contribution of the
free half-bound state.

Eq.~(\ref{unrenorm}) is divergent (the phase shifts fall as
$\frac{1}{k}$ for $k \to\infty$), which is what we should expect since it
includes the divergent contribution from the tadpole graph without the
divergent contribution from the counterterms that cancels it.

To avoid infrared problems later, we first use Levinson's theorem to
compute the change in particle number, which is given by
\begin{equation}
0 = \sum_j 1 + \int_{0}^{\infty} \frac{dk}{\pi}
\frac{d}{dk}(\delta_{\rm A}(k) + \delta_{\rm S}(k)) - \half
\end{equation}
where again, half-bound states are counted with a $\half$ in
the sum over $j$ and the $-\half$ comes from the contribution of the
free half-bound state.  Subtracting $\frac{m}{2}$ times this equation from
eq.~(\ref{unrenorm}), we  have
\begin{equation}
\Delta E = \half\sum_j (E_j-m) + \int_{0}^{\infty} \frac{dk}{2\pi}
(\omega(k)-m) \frac{d}{dk}(\delta_{\rm A}(k) + \delta_{\rm S}(k)).
\label{unrenormprime}
\end{equation}

Next, we and subtract the first Born approximation to
eq.~(\ref{unrenormprime}), which corresponds exactly to the tadpole graph. 
We must then add it back in using ordinary renormalized perturbation
theory.  However, in 1+1 dimensions we can adopt the simple renormalization
condition that the counterterms cancel the tadpole graph and perform no
additional finite renormalizations beyond this cancellation.  With this
choice, there is then nothing to add back in.

The first Born approximation is given by
\begin{eqnarray}
\delta^{(1)}_{\rm S}(k) &=& -\frac{1}{k}\int_{0}^{\infty} V(x) \cos^2 kx
\, dx
\nonumber \\
\delta^{(1)}_{\rm A}(k) &=& -\frac{1}{k}\int_{0}^{\infty} V(x) \sin^2 kx
\, dx.
\label{Borneqn}
\end{eqnarray}
Notice that the sum of these two depends on $V(x)$ only
through the quantity $\langle V\rangle = \int_{0}^{\infty} dx V(x)$, so we
can indeed cancel the tadpole contribution with available counterterms.

Subtracting the first Born approximation, we have
\begin{equation}
{\cal E}[\phi_0] = \half\sum_j (E_j-m) + \int_{0}^{\infty} \frac{dk}{2\pi}
(\omega(k)-m) \frac{d}{dk}(\delta_{\rm A}(k) + \delta_{\rm S}(k) +
\frac{\langle V\rangle}{k}).
\label{renorm1}
\end{equation}
Since the Born approximation becomes exact at large $k$, the
$\frac{\langle V \rangle}{k}$ term exactly cancels the leading
$\frac{1}{k}$ behavior of the phase shift at large $k$.  As a result,
this integral is completely finite and well-defined, since the
integrand goes like $\frac{1}{k^2}$ for $k\to\infty$ and goes to a
constant at $k = 0$.  We must find such a result, since in 1+1
dimensions eliminating the tadpole is sufficient to render the theory finite.  

In particular, we are free to integrate by parts,
giving an expression that will be easier to deal with computationally,
\begin{equation}
{\cal E}[\phi_0] = \half\sum_j (E_j-m) - \int_{0}^{\infty} \frac{dk}{2\pi}
\frac{k}{\omega(k)} (\delta_{\rm A}(k) + \delta_{\rm S}(k) +
\frac{\langle V\rangle}{k}).
\label{renorm2}
\end{equation}

Our use of Levinson's theorem to regularize
$\Delta E$ in the infrared (replacing eq.~(\ref{unrenorm}) by
eq.~(\ref{unrenormprime})) avoided a subtlety of the Born approximation in
one dimension that does not occur in higher dimensions:  as $k\to 0$, the
symmetric contribution from eq.~(\ref{Borneqn}) introduces a spurious
infrared divergence, since it goes like $\frac{1}{k}$.  (Each higher
dimension adds a power of $k$ near $k=0$.  We can see this in the Feynman
diagram calculation, where these powers of $k$ come from the measure.)

\section{Applications}

We can now use eq.~(\ref{renorm2}) to calculate quantum energies for specific
field configurations.  Knowing that our model has a ``kink''
soliton solution $\phi_0(x) = \tanh\frac{mx}{2}$, we will consider a
family of field configurations that continuously interpolate between the
trivial vacuum and a soliton/antisoliton pair,
\begin{equation}
\phi_0(x,x_0) = \tanh\frac{m}{2}(x+x_0) - \tanh\frac{m}{2}(x-x_0) - 1,
\end{equation}
with $2x_0$ measuring approximately the separation
between the soliton and antisoliton. Unlike the kink, these configurations
are not solutions of the equations of motion, except in the $x_0\to 0$ and
$x_0\to \infty$ limits.  Thus we will need to to introduce a source that
will vanish in these limits, and we must analyze the stability questions we
raised earlier.

For $x_0$ very small, we simply have a small attractive perturbation from
the trivial vacuum held in place by a small source, which we would not
expect to introduce any instabilities.  In terms of the scattering
problem, for small $x_0$, the potential is too weak to bind a state with
a binding energy greater than $m$, which would give an imaginary
eigenvalue.

For $x_0$ very large, we have a widely separated
soliton/antisoliton pair.  We know from translation invariance that a
single soliton has a mode with $\omega^2 = 0$.  Since this zero mode
corresponds to a nodeless wavefunction, it is the lowest energy mode.  The
soliton/antisoliton pair has two translation modes that will mix,
giving a symmetric eigenstate with a slightly lower energy and an
antisymmetric eigenstate with a slightly higher energy.  Thus we expect to
find a single symmetric mode with $\omega^2 < 0$, which will contribute an
imaginary part to ${\cal E}[\phi_0]$.  The imaginary part gives the rate
for our field configuration to decay through this mode toward the trivial
vacuum.  As $x_0\to\infty$, we should find that the imaginary part goes to
zero, and the real part goes to twice the energy of a single soliton, which
we compute exactly below. 

According to this analysis, there should be a finite, nonzero value of
$x_0$ where the imaginary eigenvalue first appears.  At this point,
the field becomes unstable with respect to small perturbations in some
direction in field space.  Therefore, the energy must have a lower
minimum that we can reach by moving in that direction.
This configuration $\psi_0(x,x_0)$ is a second stationary point of the
action with the same source, which crosses $\phi_0$ at the value of
$x_0$ where the imaginary part for the energy appears.
(For smaller values of $x_0$, this solution still exists but has
higher energy.)  In terms of the scattering problem, this crossing
appears when the potential has a bound state with $\omega^2 = 0$.  We
can identify this state explicitly:  since $\phi_0$ and $\psi_0$
satisfy eq.~(\ref{Jeq}) with the same $J$, as we approach the
crossing, the wavefunction $\eta(x) = \psi_0(x,x_0)-\phi_0(x,x_0)$
becomes a solution to eq.~(\ref{etaeq}) with $\omega^2 = 0$.

We have carried out this computation numerically, and find results that
agree with all of these expectations.  To compute the antisymmetric phase
shift, we will parameterize the wavefunction as
\begin{equation}
\eta(x) = e^{-ikx} - e^{ikx} e^{2i\beta(x)}
\end{equation} where $\beta(x)$ is an arbitrary complex function.  Solving
the differential equation for $\eta(x)$ subject to $\eta(0) = 0$, we find
\begin{equation}
\delta_{\rm A}(k) = - 2 \: {\rm Re} \: \beta(k,0)
\end{equation}
where $\beta(k,x)$ is 0 at $x=\infty$ and satisfies
\begin{equation}
-i\beta'' + 2 k\beta' + 2 (\beta')^2 + \half V(x) = 0,
\end{equation}
with prime denoting differentiation with respect to $x$.  To obtain the
symmetric phase shift, we can follow the same derivation, but instead of
imposing $\eta(0) = 0$ on the wavefunction, we instead impose $\eta'(0) = 0$,
giving
\begin{equation}
e^{2i\delta_{\rm S}} = \frac{e^{2i\beta}}{e^{-2i\beta^\ast}}
\frac{k+2\beta'}{k+2\beta'^\ast}
\end{equation}
with $\beta$ the same as above, so that
\begin{equation}
\delta_{\rm S}(k) = \delta_{\rm A}(k) -
\tan^{-1}(\frac{2 \: {\rm Im} \: \beta'(k,0)}{k + 2\: {\rm Re} \: \beta'(k,0)}).
\end{equation}

\begin{figure}[htbp]
$$
\BoxedEPSF{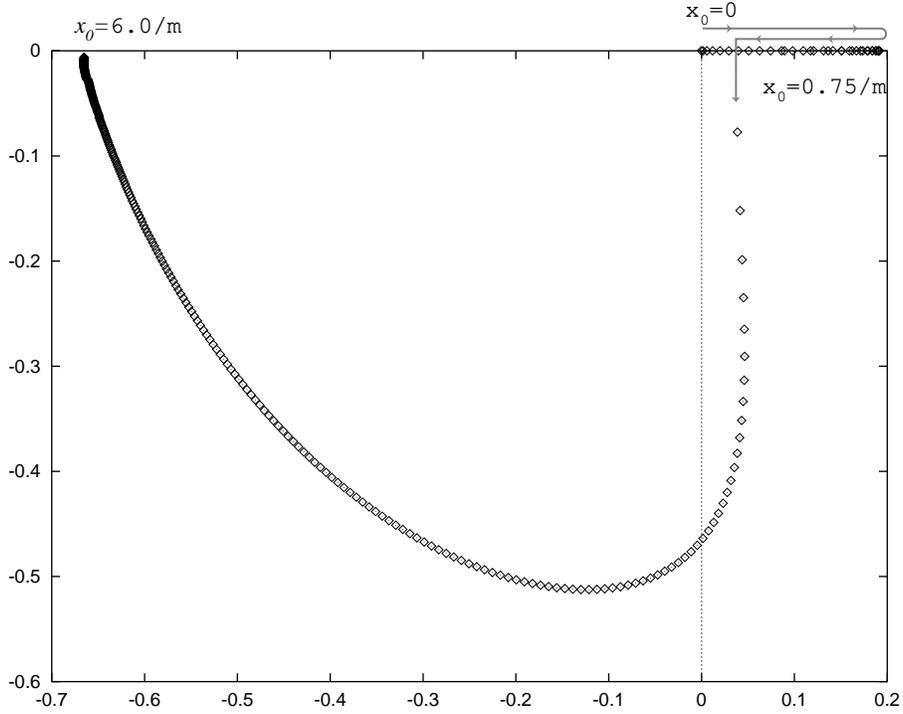 scaled 500}  
$$
\caption{The trajectory of ${\cal E}[\phi_0]$
in the complex energy plane, with ${\cal E}$ in units of
$m$, for $x_0$ increasing from $0$ to $\frac{6}{m}$ in steps of
$\frac{0.02}{m}$.  Arrows indicate the flow along the real axis.} 
\label{figure1}
\end{figure}

Fig.~\ref{figure1} shows the trajectory of ${\cal E}[\phi_0]$ in the
complex plane as a parametric function of $x_0$, starting from the
origin at $x_0 = 0$.  When $x_0\approx \frac{0.75}{m}$, a single
imaginary eigenvalue appears and ${\cal E}[\phi_0]$ leaves the real
axis.  For $x_0$ large, ${\cal E}[\phi_0]$ approaches $2(\frac{1}{4
\sqrt{3}} - \frac{3}{2\pi})$, twice the standard result for a single
soliton.  The actual trajectory of ${\cal E}[\phi_0]$ has little
significance, since it depends in detail on the functional form of
$\phi(x,x_0)$.  However, the general features --- beginning at the
origin, moving up the real axis, out into the complex plane, and
finally asymptotically to the real two-solution value --- are
characteristics of any $\phi_0$ that begins at the vacuum and ends at
a well separated kink and antikink.

\begin{figure}[htb]
$$
\BoxedEPSF{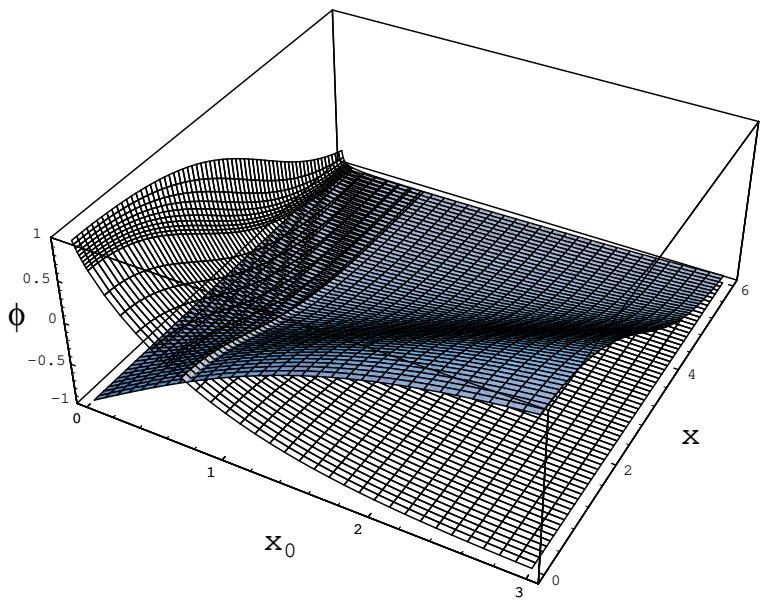 scaled 600}  
$$
\caption{Solutions to the equations of motion with a source given by
eq.~(\ref{Jeq}), as functions of $x$ and $x_0$ in units of $\frac{1}{m}$.  
The shaded graph is $\phi_0(x,x_0)$, which is guaranteed to be a
solution by the construction of $J$.  The unshaded graph gives the
second solution $\psi_0(x,x_0)$.  (A finer mesh is used for this graph
between $x_0 = 0$ and $x_0 = \frac{0.1}{m}$ in order to illustrate its
behavior in this region.) 
}
\label{figure2}
\end{figure}

Fig.~\ref{figure2} shows $\phi_0(x,x_0)$ and the second solution to
the equations of motion with the same $J$, $\psi_0(x,x_0)$.  $\psi_0$
goes to the trivial vacuum as $x_0\to\infty$, and becomes a widely
separated soliton/antisoliton pair as $x_0\to 0$, crossing $\phi_0$ at
$x_0 \approx 0.75$.  For $x_0$ below this value, $\phi_0$ has lower
classical energy, while above this point, $\psi_0$ has lower classical
energy, and this crossing appears precisely where the imaginary
eigenvalue appears in the small oscillations spectrum.

Through this continuous deformation from the trivial vacuum, we have
arrived at a widely separated soliton/antisoliton pair, which we can
now separate into independent configurations with nontrivial
topology.  These configurations are exactly soluble, so we will be
able to study them analytically.  However, our method does not rely on
having an analytic solution, so we could numerically calculate the energy of a
generic field configuration with nontrivial topology using the same techniques.

\section{Analytic results}

In one dimensional quantum mechanics, potentials of the form
\begin{equation}
V_\ell(x) = -\frac{\ell+1}{\ell}m^2{\rm sech}^2(\frac{mx}{\ell})
\end{equation}
with $\ell$ an integer are exactly soluble and reflectionless.  
Their properties are summarized in  Appendix A.  The single
soliton solution in our model, $\phi_0(x) = \tanh\frac{mx}{2}$,
corresponds to $V(x) = -\frac{3}{2}m^2{\rm sech}^2\frac{mx}{2}$, the
$\ell=2$ case of this family. (The sine-Gordon soliton corresponds to the
$\ell=1$ case.) For a reflectionless potential, $\delta_{\rm S}(k) =
\delta_{\rm A}(k)$; to reconcile this equality with Levinson's theorem in
the symmetric and antisymmetric channels, $V_l(x)$ must have a half-bound
state.  (We saw this behavior already for the $\ell = 0$ case, the free
particle.)  We also note that although  $\delta_{\rm S} = \delta_{\rm A}$, 
$\delta^{(1)}_{\rm S} \neq \delta^{(1)}_{\rm A}$, so the renormalized
contributions from the symmetric and antisymmetric channels are not the
same.  In addition, the bound state contributions will also be unequal.

For our  case, the exact result for the phase shift is
\begin{equation}
\delta_{\rm S}(k) = \delta_{\rm A}(k) =  -\tan^{-1} \frac{3mk}{m^2-2k^2},
\end{equation}
where the branch of the arctangent is chosen so that the phase shift is
continuous and goes to zero for $k\to\infty$.  The Born approximation is
\begin{equation}
\delta^{(1)}_{\rm S}(k) + \delta^{(1)}_{\rm A}(k) = \frac{3m}{k}.
\end{equation}

There are three bound states: a translation mode with $E=0$, a state with
$E = \frac{m\sqrt{3}}{2}$, and a half-bound state with $E=m$.
Using eq.~(\ref{renorm1}) or eq.~(\ref{renorm2}) we find
\begin{equation}
{\cal E}[\phi_0] = m(\frac{1}{4\sqrt{3}} - \frac{3}{2\pi})
\end{equation}
in agreement with \cite{DHN}, \cite{Bordag}, and the
mode number cutoff method in \cite{vN1}.

We can calculate the energy of the soliton of the sine-Gordon model using
the same methods.  In this model, the $\frac{m^2}{4}(\phi^2 - 1)^2$
potential is replaced by
\begin{equation}
U(\phi) = m^2(\cos\phi - 1)
\end{equation}
which has soliton (anti-soliton) solutions
\begin{equation}
\phi_0 = 4 \tan^{-1}(\exp(\mp m(x-x_0))).
\end{equation}
The phase shift is
\begin{equation}
\delta_{\rm S}(k) = \delta_{\rm A}(k) =  \tan^{-1} \frac{m}{k}
\end{equation}
with Born approximation
\begin{equation}
\delta^{(1)}_{\rm S}(k) + \delta^{(1)}_{\rm A}(k) =  \frac{2m}{k}.
\end{equation}
The bound states are just the translation mode at $E=0$ and the
half-bound state at $E=m$, giving
\begin{equation}
{\cal E}[\phi_0] = -\frac{m}{\pi}
\end{equation}
again agreeing with the established results.

\clearpage
\newpage

\chapter{Fermions in one dimension}

We next extend our methods by adding fermions.  We will see that our
techniques extend cleanly and unambiguously to this case, and deal
elegantly with the subtleties of boundary conditions.

\section{Formalism}

We consider a Majorana fermion $\Psi$ interacting with a scalar background
field $\phi$, with the classical Lagrangian density
\begin{equation}
  {\cal L} = \frac{m^2}{2\lambda} \left(
    i\bar\Psi \slash\hspace{-0.5em}\partial \Psi - m\phi \bar \Psi
    \Psi + {\cal L}_\phi \right)
\end{equation} where ${\cal L}_\phi$ is the Lagrangian density for the
$\phi$ background field, which we will take to be the same as in the last
section, with the same soliton solutions.  We note that choosing the
Lagrangian density in this way causes the bosonic and fermionic degrees of
freedom to be related by supersymmetry for a background field that is a
solution to the equations of motion (such as a single soliton or an
infinitely separated soliton/antisoliton pair).  In this section, we will
use this property only because we will find it instructive to compare
the bosonic and fermionic small oscillations spectra.  In the next
section, we will consider the full supersymmetric model.

The one-loop corrections to the energy due to fermionic fluctuations
will be given by the appropriately renormalized sum of the zero-point
energies, $-\half\omega$, of the fermionic small oscillations.  The
spectrum of fermionic small oscillations in a background $\phi_0(x)$
is given by the Dirac equation,
\begin{equation}
  \gamma^0 \left(-i\gamma^1 \frac{d}{dx} + V_F(x)\right) \psi_k(x) =
\omega^F_k
  \psi_k(x)
  \label{Diraceq}
\end{equation}
where $V_F(x) = m\phi_0(x)$ is the fermionic potential and
$k=\pm\sqrt{\omega^{2}-m^{2}}$ is the momentum labeling the scattering
states.

We will choose the convention $\gamma^0 = \sigma_2$ and $\gamma^1 = i
\sigma_3$ so that the Majorana condition becomes simply $\Psi^\ast = \Psi$.
We note that since our Lagrangian is $CP$ invariant, all of our
results for the spectrum of a Majorana fermion can be extended to a
Dirac fermion simply by doubling.

Again, we will need to find the phase shifts.  We can solve for the phase
shifts of any fermionic potential $V_F(x)$ that satisfies $V_F(x) =
V_F(-x)$ and $V_F(x)\to m$ as $x\to\pm\infty$ by generalizing the methods
of the previous section.  This form will be useful for
considering our example of a sequence of background field configurations
that continuously interpolates between the trivial vacuum and a widely
separated soliton-antisoliton pair. In the limit of infinite separation,
the phase shift for the pair goes to twice the result for a single
soliton.  For comparison, we also do the computation for a single soliton
directly below.

We define the parity operator acting on fermionic states as $P=\Pi
\gamma^0$, where $\Pi$ sends $x\to -x$.  $P$ commutes with the
Hamiltonian, so parity is a good quantum number.  As a result, we can
separate the small oscillations into positive and negative parity
channels, now restricted to $k>0$ in the continuum.  

We parameterize the fermion solutions by
\begin{equation}
\chi_1(x) = \left( \matrix{
        e^{i\nu(x)} \cr
        i e^{i\zeta(x)} e^{i\theta}
} \right) e^{ikx}
\hbox{~~~~and~~~~}
\chi_2(x) = \left( \matrix{
        e^{-i\nu(x)^\ast} \cr
        i e^{-i\zeta(x)^\ast} e^{-i\theta}
} \right) e^{-ikx}
\end{equation}
where $\theta = \tan^{-1} \frac{k}{m}$ and $\nu$ and $\zeta$ are
complex functions of $x$.  We then find the phase shift in each
channel $\delta^{\pm}(k)$ by solving eq.~(\ref{Diraceq}) subject to
the boundary conditions
\begin{equation}
\psi^{\pm}(0) \propto \left( \matrix{ 1 \cr \pm i } \right)
\end{equation}
giving the phase shift as 
\begin{equation}
\psi^{\pm}(0) = e^{\frac{i\theta}{2}} \chi_2(0) \pm
 e^{2i\delta^\pm(k)} e^{-\frac{i\theta}{2}} \chi_1(0).
\end{equation}
Our boundary conditions assure that the wavefunctions $\psi^{\pm}$
are eigenstates of the parity operator with eigenvalues $\pm 1$.  
We obtain
\begin{eqnarray}
\delta^+(k) &=& -{\rm Re} \: \nu(0) + \frac{\theta}{2} + \frac{1}{2i}
\log \frac{Y-1}{1-Y^\ast} \cr
\delta^-(k) &=& -{\rm Re} \: \nu(0) + \frac{\theta}{2} + \frac{1}{2i}
\log \frac{1+Y}{1+Y^\ast}
\end{eqnarray}
where $Y = \frac{1}{\omega}(V_F(0) - ik + i\nu^\prime(0)^\ast)$
and $\nu(x)$ satisfies
\begin{equation}
-i\nu(x)^{\prime\prime} + \nu^\prime(x)^2 + 2k\nu^\prime(x) +
V_F(x)^2 - V_F(x)^\prime - m^2 = 0
\label{nueq}
\end{equation}
with the boundary condition that $\nu(x)$ and $\nu^\prime(x)$ vanish
at infinity.  The total phase shift is given by summing the phase shifts
in each channel, $\delta_F(k) = \delta^+(k) + \delta^-(k)$.

Again, once we know the phase shifts, Levinson's theorem tells us how
many bound states there will be.  It works exactly the same way as in the
bosonic case \cite{lev}: In the odd parity channel, the number of bound
states $n_-$ is given by
\begin{equation}
\delta^-(0) = \pi n_-
\label{Lev1}
\end{equation}
while in the even parity channel the number of bound states $n_+$ is given
by
\begin{equation}
\delta^+(0) = \pi (n_+ - \half).
\label{Lev2}
\end{equation}
One can derive this result either by the same Jost function
methods used in the boson case, or by observing that at small $k$, the
nonrelativistic approximation becomes valid so the bosonic results carry
over directly.  As in the boson case, for a particular potential there may
exist a $k=0$ state in either of the two channels whose Dirac wavefunction
goes to a constant spinor as $x\to\pm\infty$. (Generically, for $k=0$ the
components of the Dirac wavefunction go to straight lines as
$x\to\pm\infty$, but not lines with zero slope.)  Just as in the bosonic
case, such threshold states (which we will again call ``half-bound
states'') should be counted with a factor of $\half$ in Levinson's theorem.

Given the phase shifts and bound state energies, we can calculate the 
one-loop fermionic correction to the energy.  We continue to work in the
simple renormalization scheme in which we add a counterterm proportional to
$\phi^2 - 1$, and perform no further renormalizations.  The counterterm is
fixed by requiring that the tadpole graph cancel.  As in the bosonic case,
we use the density of states
\begin{equation}
\rho(k) = \rho_0(k) + \frac{1}{\pi} \frac{d\delta_F(k)}{dk}
\end{equation}
to write
\begin{equation}
{\cal E}[\phi_0] = -\half \sum_j \omega_j - \int_0^\infty \frac{dk}{2\pi}
\omega \frac{d\delta_F(k)}{dk}  + {\cal E}_{\rm ct}
\end{equation}
where $\omega = \sqrt{k^2+m^2}$  and the sum over $j$ counts bound states
with appropriate factors of $\half$ as discussed above.  Next, we use
eq.~(\ref{Lev1}) and eq.~(\ref{Lev2}) to rewrite this expression as
\begin{equation}
{\cal E}[\phi_0] = -\half \sum_j (\omega_j-m) - \int_0^\infty \frac{dk}{2\pi}
(\omega-m) \frac{d\delta_F(k)}{dk} + {\cal E}_{\rm ct}.
\end{equation}
In terms of the shifted field $\phi - 1$, the (divergent) contribution
from the tadpole graph is given by the leading Born approximation to
$\delta_F$.  A corresponding divergence related by $\phi \to -\phi$
symmetry appears in the second-order diagram.  To subtract the tadpole
graph, we subtract the leading Born approximation in $V_F - m$.  In
order to maintain the $\phi\to-\phi$ symmetry, we also subtract the
corresponding piece of the second-order diagram by subtracting the
part of the second Born approximation related by the symmetry.
Since we have chosen the counterterm to exactly cancel these
subtractions, there is nothing to add back in.  Thus we have
\begin{equation}
{\cal E}[\phi_0] = -\half \sum_j (\omega_j -m) - \int_0^\infty \frac{dk}{2\pi}
(\omega - m) \frac{d}{dk} (\delta_F(k) - \delta_F^{(1)}(k))
\label{totenergy}
\end{equation}
where
\begin{equation}
\delta^{(1)}(k) = -\frac{1}{k} \int \left(V_F(x)^2 - V_F(x)^\prime -
  m^2 \right) \, dx
\end{equation}
which can also be obtained numerically by iterating
eq.~(\ref{nueq}). For the soliton/antisoliton pair, the contribution from
$V^\prime(x)$ vanishes since it is a total derivative, so that again our
subtraction is indeed proportional to $\phi^2-1$.

\section{Applications}

We continue to study a sequence of background fields labeled by a parameter
$x_0$ that continuously interpolates from the trivial vacuum $\phi(x)=1$ at
$x_{0}=0$ to a widely separated soliton/antisoliton pair at
$x_{0}\to\infty$,
\begin{equation}
\phi(x,x_0) = 1 + \tanh \frac{m(x-x_0)}{2} - \tanh \frac{m(x+x_0)}{2}.
\end{equation}
\begin{figure}[htbp]
$$
\BoxedEPSF{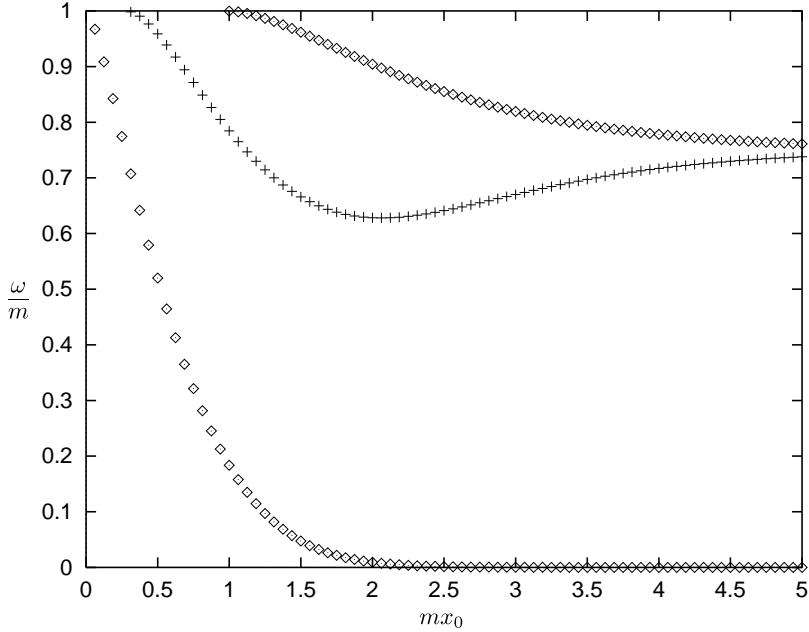 scaled 600}  
$$
\caption{Fermion bound state energies as a function of $mx_0$ in units
of $m$.  Even parity states are denoted with $\diamondsuit
\hspace{-0.525em} \cdot$~~and odd parity states with $+$.}
\label{boundf}
\end{figure}

Fig.~\ref{boundf} shows the fermionic bound state energies as a function of
$x_0$.  In the limit $x_0\to\infty$ two bound states approach energy $m
\frac{\sqrt{3}}{2}$.  These are  simply the odd and even parity
combinations of the single soliton bound state at $m \frac{\sqrt{3}}{2}$.
The third (positive parity) bound state approaches $\omega = 0$ where the
single soliton also has a bound state, but there is only {\it one} such
state (with $\omega >0$).  Thus for a single soliton we must thus count the
zero mode with a factor of $\half$.  We will see this result analytically
below. 

For any finite $x_{0}$, Levinson's theorem holds
without subtleties; there  are no states that require factors of
$\half$.  Thus for a large but finite $x_{0}$, we find
\begin{eqnarray}
\delta^+(0) &=& \frac{3\pi}{2} \cr \delta^-(0) &=& \pi
\end{eqnarray}
consistent with having two positive parity and one negative parity
bound states (see Fig.~\ref{boundf}).  In the limit $x_0\to\infty$,
an even parity ``half-bound'' threshold state enters the spectrum at
$\omega = m$  just as in the bosonic case.  Also, in this
limit, the lowest (positive parity) mode approaches $\omega =
0$, and is only counted as a $\half$ as described above.
Finally, a negative parity mode enters the spectrum
from below, also to be weighted by $\half$.

\begin{figure}[htbp]
$$
\BoxedEPSF{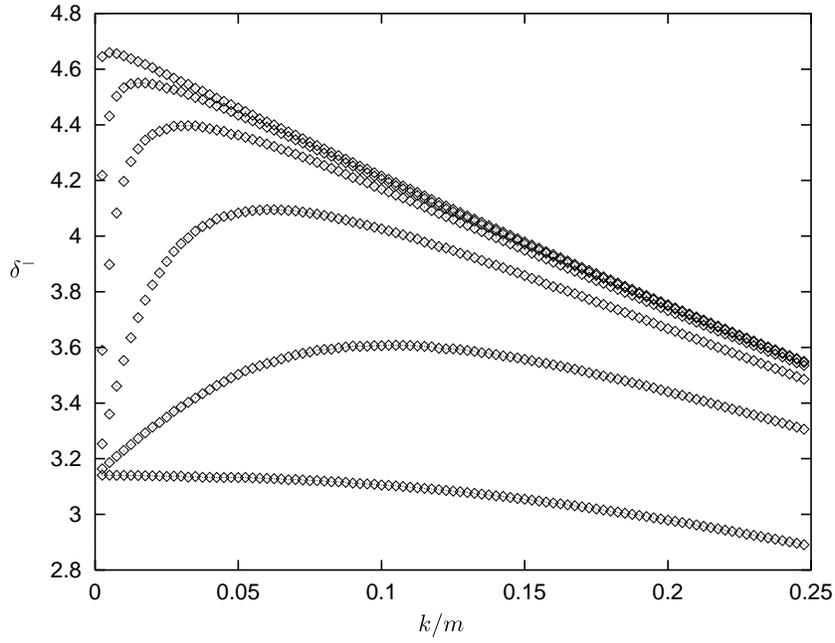 scaled 600}  
$$
\caption{Negative-parity phase shifts as functions of $k/m$ for $x_0 =
$ 2.0, 3.0, 4.0, 5.0, 6.0, and 8.0.  For any finite separation, the
phase shift is equal to $\pi$ at $k=0$, but as $x_0$ gets larger, the
phase shift ascends more and more steeply to $\frac{3\pi}{2}$.
}
\label{phasefa}
\end{figure}

Thus the net change is to add half of a negative parity
state, which via Levinson's theorem requires the phase shift
$\delta^{-}(0)$ to jump from $\pi$ to $\frac{3\pi}{2}$
as $x_{0}\to\infty$.  This jump occurs by the same continuous
but nonuniform process as in all cases where a new state gets bound,
which is illustrated in Fig.~\ref{phasefa}.  Just as in the bosonic
case, in the limit of infinite separation the potential we have chosen
becomes reflectionless, which requires $\delta^+(k) = \delta^-(k)$.

\begin{figure}[htbp]
$$
\BoxedEPSF{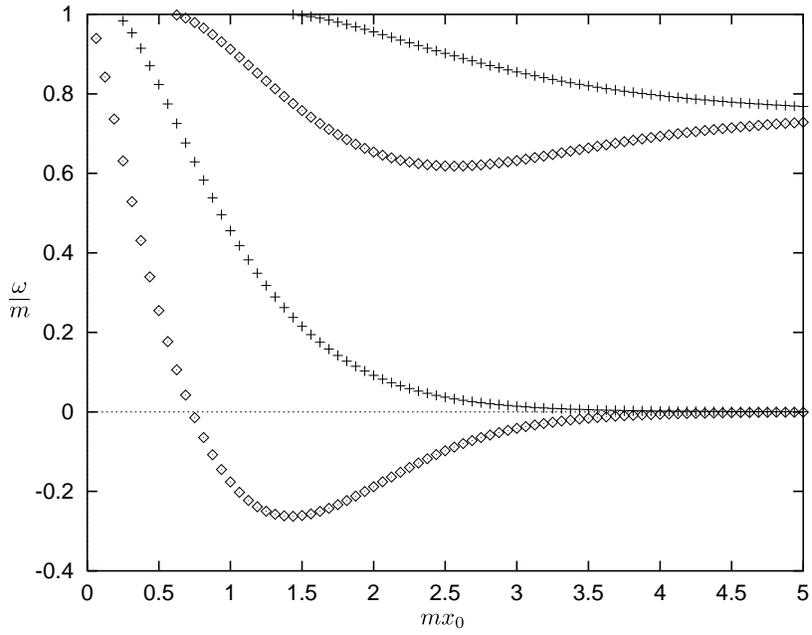 scaled 600}  
$$
\caption{Boson bound state squared energies as a function of $mx_0$ in
units of $m$.  Symmetric states are denoted with $\diamondsuit
\hspace{-0.525em} \cdot$~~and antisymmetric states with $+$.}
\label{boundb}
\end{figure}

To contrast the behavior of the zero modes, Fig.~\ref{boundb} shows 
$\omega^2$ for the bound states of the bosonic small oscillations.
Because we have chosen the bosonic potential to be 
consistent with supersymmetry, the bosonic and fermionic spectra are 
related.  Again as $x_0 \to\infty$ the bound state energies approach 
those of the single soliton, and the wavefunctions are formed from the 
odd and even combinations of the wavefunctions for the single soliton.  
However, we see that in the boson case both the mode at $m 
\frac{\sqrt{3}}{2}$ and the mode at $\omega = 0$ are doubled, so there
is no factor of $\half$ in counting the bosonic zero modes.

\begin{figure}[htbp]
$$
\BoxedEPSF{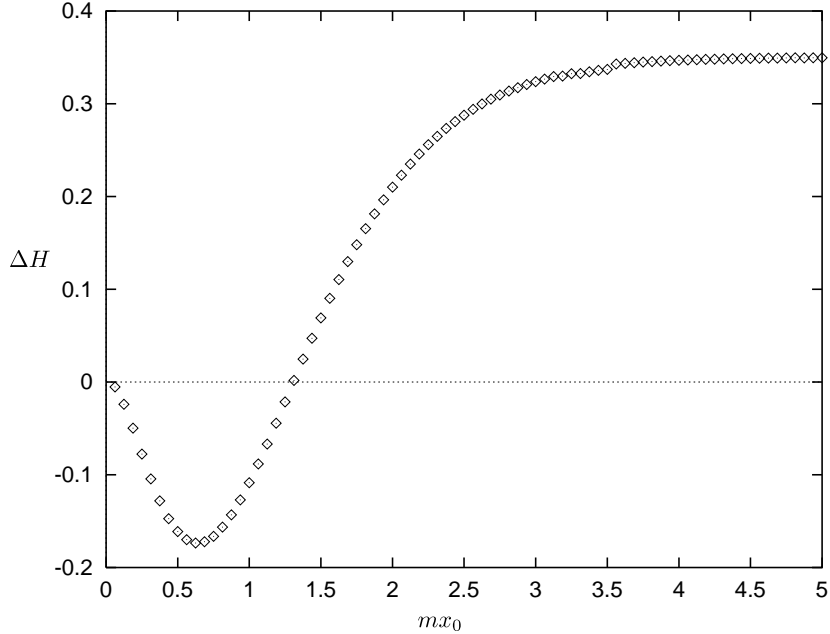 scaled 600}  
$$
\caption{One-loop fermionic correction to the energy as a function of
  $x_0$ in units of $m$.}
\label{energyf}
\end{figure} Fig.~\ref{energyf} shows the values of ${\cal E}[\phi_0]$ we
obtain from eq.~(\ref{totenergy}) as a function of $x_0$.  In the
$x_0\to\infty$ limit, we can also do the calculation analytically.  We will
expand on this technique in our discussion of the supersymmetric case
below.  We find a phase shift
\begin{equation}
\delta_F(k) = 4\tan^{-1}\frac{m}{2k} + 2\tan^{-1}\frac{m}{k}
\end{equation}
with Born approximation
\begin{equation}
\delta_F^{(1)}(k) = \frac{4m}{k}
\end{equation}
and contributions from two bound states at $\omega =
m\frac{\sqrt{3}}{2}$ and one at $\omega = m$.  We thus find
\begin{equation}
{\cal E}[\phi_0] = 2m \left(\frac{1}{\pi} - \frac{1}{4 \sqrt{3}}\right)
\end{equation}
which agrees with the numerical calculation.

\clearpage
\newpage

\chapter{The supersymmetric model in one dimension}

Having analyzed bosons and fermions separately, we now combine these results
and consider the full supersymmetric mode in one dimension.  Here we
will work analytically using a general potential, with the $\phi^4$ and
sine-Gordon theories providing two concrete examples.  The
supersymmetric model also introduces another quantity, the central
charge, which we must also compute in order to check the consistency
of our results for the energy.  A review of the literature
shows a wide variety of conflicting results\cite{vN1,many,KR,vN2,Sch,IM}.
Our techniques for dealing with the  regularization, renormalization
and calculation of one-loop corrections to the energies of classical
field configurations will enable us to take a fresh look at this problem.  
We will obtain an unambiguous result for the energy and the central
charge, and in the process we will see how our techniques are related
to the expansion of the quantum field in creation and annihilation operators.

\section{Formalism}

We begin with the classical supersymmetric Lagrangian density in 1+1 dimensions
\begin{equation}
  {\cal L} = \frac{m^2}{2\lambda} \left((\partial_\mu \phi)
    (\partial^\mu \phi) - U(\phi)^2 +i\bar\Psi 
    \slash\hspace{-0.5em}\partial \Psi - U^\prime(\phi) \bar \Psi 
    \Psi \right)
  \label{LaG}
\end{equation}
where $\phi$ is a real scalar, $\Psi$ is a Majorana fermion, and
$U(\phi) = \frac{m}{2}(\phi^2 - 1)$  for the $\phi^4$ model and
$U(\phi) = 2m \sin(\phi/2)$ for the sine-Gordon model.
These models support classical soliton and antisoliton 
solutions, which are the solutions to
\begin{equation}
  \phi_0^\prime(x) = \mp U(\phi_0)
  \label{virial}
\end{equation}
for the soliton and antisoliton respectively.  In the $\phi^4$ model,
the soliton solution is the ``kink'' that we have already discussed,
\begin{equation}
\phi_0^{\rm kink}(x) = \tanh\frac{mx}{2},
\end{equation}
while the soliton in the sine-Gordon model is given by
\begin{equation}
\phi_0^{\rm SG}(x) = 4 \tan^{-1} e^{-mx},
\end{equation}
and in both cases the antisoliton is obtained from the soliton by
sending $x\to -x$.  

The eigenvalue equations for the bosonic and fermionic small
oscillations are
\begin{eqnarray}
  \left( -\frac{d^2}{dx^2} + U^\prime(\phi_0)^2 + 
    U(\phi_0)U^{\prime\prime}(\phi_0) \right) \eta_k(x) &=& 
  (\omega_k^B)^2 \eta_k(x) \\
  \gamma^0 \left(-i\gamma^1 
    \frac{d}{dx} + U^\prime(\phi_0)\right) \psi_k(x) &=& \omega^F_k 
  \psi_k(x).
  \label{DiracKGeq}
\end{eqnarray}

The bosonic potentials are given by 
\begin{equation}
  U^\prime(\phi_0)^2 + 
  U(\phi_0)U^{\prime\prime}(\phi_0) - m^2 = 
  -\left(\frac{\ell+1}{\ell} \right) m^2
  {\rm sech}^2\frac{mx}{\ell} \equiv V_\ell(x) 
  \label{Vl}
\end{equation}
with $\ell=1$ for the sine-Gordon soliton and $\ell=2$ for the kink.  
Repeating our results from above, we have phase shifts
\begin{eqnarray}
  \delta_B^{\rm kink}(k) = 
  \delta_{\ell=2}(k) &=& 2\tan^{-1}\left(\frac{3mk}{2k^2-m^2}\right) \cr
  \delta_B^{\rm SG}(k)= \delta_{\ell=1}(k) &=& 2\tan^{-1}\frac{m}{k}
\end{eqnarray}
and bound  states at $\omega=0$ for the kink
$\omega=\frac{\sqrt{3}m}{2}$, and at $\omega=0$ for the sine-Gordon
soliton.  As discussed above, both potentials have half-bound states
at threshold, a necessary consequence of their reflectionless property.
  
As above, we avoid subtleties of boundary conditions by computing the
fermionic phase shifts for a widely separated
soliton and antisoliton solution, with the antisoliton on the 
left so that $U^\prime(\phi) \to m$ as 
$x\to\pm\infty$.  We will use the second-order equation obtained by 
squaring eq.~(\ref{DiracKGeq}),
\begin{equation}
  \left( \matrix{ -\frac{d^2}{dx^2} + V_\ell(x) & 0 \cr 0 & 
      -\frac{d^2}{dx^2} + \tilde V_\ell(x) } \right) \psi_k(x) = k^2 
  \psi_k(x)
\label{secondordersol}
\end{equation}
for the soliton and
\begin{equation}
  \left( \matrix{ -\frac{d^2}{dx^2} + \tilde V_\ell(x) & 0 \cr 0 
      & -\frac{d^2}{dx^2} + V_\ell(x)\cr } \right) \psi_k(x) = k^2 
  \psi_k(x)
\end{equation}
for the antisoliton, where $\tilde V_\ell(x) = \frac{1}{\ell^2} 
V_{\ell-1}(\frac{x}{\ell})$.  An incident wave far to the left is 
given by
\begin{equation}
  \psi_k(x) = e^{ikx} \left( \matrix{ 1 \cr i e^{i\theta} } 
  \right)
\end{equation}
with $\theta = \tan^{-1}\frac{k}{m}$.  To start with, we will restrict
to the case of a reflectionless potential, such as the kink or sine-Gordon
soliton.  It thus scatters without reflection  through the antisoliton becoming
\begin{equation}
  \psi_k(x) = e^{ikx} \left( \matrix{ e^{i\tilde\delta_\ell} \cr i 
      e^{i(\delta_\ell + \theta)} } \right)
\label{inbetween}
\end{equation}
where $\delta_\ell(k)$ is the phase shift for the bosonic potential 
$V_\ell(x)$ and $\tilde\delta_\ell(k)$ is the phase shift for the 
bosonic potential $\tilde V_{\ell}(x)$.  It then scatters without 
reflection through the soliton giving
\begin{equation}
  \psi_k(x) = e^{ikx} \left( \matrix{ e^{i(\tilde\delta_\ell + 
        \delta_\ell)} \cr i e^{i(\tilde\delta_\ell + \delta_\ell + 
        \theta)} } \right).
\end{equation}
By rescaling the results from Appendix A, we easily obtain
\begin{equation}
  \tilde \delta_\ell(k) = \delta_\ell(k) - 2\tan^{-1}\frac{m}{k}
\end{equation}
so that for the soliton/antisoliton pair,
\begin{equation}
  \delta_F(k) = \delta_B(k) - 2\tan^{-1}\frac{m}{k} 
  \label{phasedeficit2}
\end{equation}
and thus for a single soliton
\begin{equation}
  \delta_F(k) = \delta_B(k) - \tan^{-1}\frac{m}{k}
  \label{phasedeficit}
\end{equation}
in both models.  This result has been also been obtained in \cite{KR}
and \cite{vN2}.  Through this analysis, we see that the deficit 
between the boson and fermion phase shifts is necessary so that 
eq.~(\ref{inbetween}) correctly solves the Dirac equation in the 
region where $V_F(x) = - m$.  This result also agrees with the methods of the 
previous section.

These results generalize to any supersymmetric potential
$U(\phi)$ that supports a soliton $\phi_0$ with $\phi_0(x) = -\phi_0(-x)$.
We can still consider eq.~(\ref{secondordersol}), with $V_\ell(x)$ and
$\tilde V_\ell(x)$ replaced by
\begin{equation}
V(x) =  U^\prime(\phi_0)^2 +U(\phi_0)U^{\prime\prime}(\phi_0) - m^2
\end{equation}
and
\begin{equation}
\tilde V(x) = U^\prime(\phi_0)^2 - U(\phi_0)U^{\prime\prime}(\phi_0) - m^2.
\end{equation}
These are symmetric, though now not necessarily reflectionless,
bosonic potentials.  We decompose their solutions into symmetric and
antisymmetric channels.  For $x\to\pm\infty$, these solutions are
given in terms of phase shifts as 
\begin{eqnarray}
\eta_k^S(x) = \cos(kx \pm \delta^S(k)) &\qquad&
\eta_k^A(x) = \sin(kx \pm \delta^A(k)) \cr
\tilde\eta_k^S(x) = i\cos(kx \pm \tilde\delta^S(k)) &\qquad&
\tilde\eta_k^A(x) = -i\sin(kx \pm \tilde\delta^A(k))
\label{generalphase}
\end{eqnarray}
where the arbitrary factors of $\pm i$ are inserted for convenience later.
For all $x$ these wavefunctions are related by
\begin{eqnarray}
  \omega_k \tilde \eta_k^S(x) = i\left( \frac{d}{dx} + U^\prime(\phi_0)
  \right) \eta_k^A(x) &\qquad&
  \omega_k \eta_k^A(x) = i\left( \frac{d}{dx} - U^\prime(\phi_0)
  \right) \tilde\eta_k^S(x) \cr
  \omega_k \tilde \eta_k^A(x) = i\left( \frac{d}{dx} + U^\prime(\phi_0)
  \right) \eta_k^S(x) &\qquad&
  \omega_k \eta_k^S(x) = i\left( \frac{d}{dx} - U^\prime(\phi_0)
  \right) \tilde\eta_k^A(x) \cr
\label{Diracrel}
\end{eqnarray}
so that the solutions to the Dirac equation are
\begin{equation}
  \psi_k^+(x) = \left( \matrix{ \eta_k^S \cr \tilde \eta_k^A }
  \right) \hbox{~~~~and~~~~}
  \psi_k^-(x) = \left( \matrix{ \eta_k^A \cr \tilde \eta_k^S } \right)
\end{equation}
with positive and negative parity respectively.
The phase relation between the upper and lower components of these
wavefunctions must be different as $x\to\pm\infty$ since the
mass term has opposite signs in these two limits.

Putting this all together gives, as $x\to\pm\infty$,
\begin{eqnarray}
\cos(kx \pm \delta^S(k)) &=& \frac{1}{\omega_k}
\left(\frac{d}{dx}-V_F(x)\right) \sin(kx \pm \tilde\delta^A(k)) \cr
&=& \frac{1}{\omega_k}\left(k \cos(kx \pm \tilde\delta^A(k)) \mp
m \sin(kx \pm \tilde\delta^A(k)) \right) \cr
&=& \mp \sin(kx \pm \tilde\delta^A(k) \mp \theta)  =
\cos(kx \pm \tilde\delta^A(k) \mp \theta \pm \frac{\pi}{2})
\end{eqnarray}
and thus
\begin{equation}
\delta^S(k) = \tilde\delta^A(k) + \tan^{-1} \frac{m}{k}
\end{equation}
and similarly
\begin{equation}
\delta^A(k) = \tilde\delta^S(k) + \tan^{-1} \frac{m}{k}.
\end{equation}
The fermion phase shift in each channel is given by the average of the
bosonic phase shifts of the two components
\begin{eqnarray}
\delta^+(k) &=& \half(\delta^S(k) + \tilde \delta^A(k)) =
\delta^S(k) -  \half \tan^{-1} \frac{m}{k} \nonumber\\
\delta^-(k) &=& \half(\delta^A(k) + \tilde \delta^S(k)) =
\delta^A(k) -  \half \tan^{-1} \frac{m}{k} \nonumber\\
\end{eqnarray}
so that
\begin{equation}
\delta_F(k) = \delta^+(k) + \delta^-(k) = \delta^S(k) + \delta^A(k) -
\tan^{-1}\frac{m}{k}
\end{equation}
and we obtain the same result, eq.~(\ref{phasedeficit}), as we found in the
reflectionless case.

Since $\delta_{F}(0)\ne\delta_{B}(0)$, Levinson's theorem requires that the 
spectrum of fermionic and boson bound states differ.  The difference
is that, although there is a fermionic bound state for every bosonic
bound state, the mode at $\omega=0$ only counts as  $\half$ for the
fermions.  (The fermionic states at threshold also count as $\half$,
the same as in the boson case.)  We have seen this difference in the
previous section by the tracking the bound state energies as we
interpolated between the trivial vacuum and a soliton/antisoliton pair.
We can also check this result  analytically by observing that the
residue of the pole at $k=im$ in the reflection coefficient ${\cal T}_{F}$
is half the residue of the pole at $k=im$ in ${\cal  T}_{B}$ because
of eq.~(\ref{phasedeficit}).  As a final check,  we imagine doubling
the spectrum by turning $\phi$  into a complex scalar and $\Psi$ into
a Dirac fermion.  Then in a soliton background $\phi$ would have two
zero-energy bound states, one involving its real part and one
involving its imaginary part. However, $\Psi$ would have only a single
zero-energy bound state, with wavefunction given by
\begin{equation}
  \psi(x) = \left( \matrix{ e^{-\int_0^x V_F(y) dy} \cr 0 } \right)
\end{equation}
with $V_F(x) = U'(\phi_0)$. The corresponding solution with only an
lower component is not normalizable; for an antisoliton, we would find
the same situation with upper and lower components reversed. Thus when we
reduce to a Majorana fermion, we count this state as a half.

We note that the fermionic phase shift and bound state spectrum are simply
given by the average of the results we would obtain for the two
bosonic potentials $V_\ell(x)$ and $\tilde V_\ell(x)$.  We also note
that just as the bosonic zero mode arises because the  soliton breaks
translation invariance, the fermionic zero mode arises as a
consequence of broken supersymmetry invariance (which we can think of
as breaking translation invariance in a fermionic direction in
superspace).  For a soliton solution, only the supersymmetry generator
$Q_-$ is broken, while $Q_+$ is left unbroken (the situation is
reversed for an antisoliton).  Thus since the supersymmetry is
only half broken, it is not surprising that the corresponding zero
mode counts only as a half.  In both cases, acting with the broken
generator on the soliton solution gives the corresponding zero mode.

\section{Applications}
To compute the one-loop correction to the energy, we now follow the same
method as in the previous sections and sum the quantity $\half \omega$
over bosonic and fermionic states, with the fermions entering with a
minus sign as usual.  We will discuss the case of an isolated soliton,
and see that results agree with the widely separated soliton/antisoliton pair
considered in the last section.

Thus our formal expression for the energy correction is 
\begin{equation}
\Delta H = \half\sum_j \omega_j^B - \half\sum_j \omega_j^F
 + \int_0^\infty \frac{dk}{2\pi} \omega \left(\frac{d \delta_B}{dk} -
 \frac{d \delta_F}{dk} \right)
\label{unreg1}
\end{equation}
where the states at threshold and the fermion bound state at $\omega=0$ 
are weighted by $\half$ as discussed above.  The free
density of states has cancelled between bosons and fermions,
as required by supersymmetry.  Again, to avoid infrared problems
later, we use Levinson's theorem 
\begin{equation}
\delta(0) = \pi(n_B - \half)
\label{Levinson}
\end{equation}
to rewrite eq.~(\ref{unreg1}) as
\begin{equation}
        \Delta H = \half\sum_j (\omega_j^B-m) - \half\sum_j 
        (\omega_j^F -m) + \int_0^\infty \frac{dk}{2\pi} (\omega-m) 
        \left(\frac{d \delta_B}{dk} - \frac{d \delta_F}{dk}\right)
        \label{unreg2}
\end{equation}
where the $\half$ in eq.~(\ref{Levinson}) has cancelled between 
bosons and fermions.

The continuum integral in eq.~(\ref{unreg2}) is still logarithmically
divergent at large $k$, as we should expect since we have not yet
included the contribution from the counterterm.  As 
above, we can isolate this divergence in the contributions from
the low order Born approximations to the phase shifts $\delta_{B}$ and
$\delta_{F}$.  We then identify these contributions with specific
Feynman graphs, subtract the Born approximations, and add back in the
associated graphs.  For the boson, the divergence comes from the first Born
approximation, which corresponds exactly to the tadpole graphs
with a bosonic loop.  For the fermion, the source of the divergence
is more complicated: we subtract the first Born approximation to the
fermionic phase shift and the piece of the second Born approximation
that is related to it by the spontaneous symmetry breaking of $\phi$.  This
subtraction corresponds exactly to subtracting the tadpole graph with
a fermionic loop and the part of the graph with two external
bosons and a fermionic loop that is related to the tadpole graph by
spontaneous symmetry breaking (the rest of the two-point
function is then finite).  For both the boson and fermion, this subtraction
amounts to simply subtracting the term proportional to $\frac{1}{k}$ that
cancels the leading $\frac{1}{k}$ behavior of the phase shift at
large $k$.  We can identify the coefficient of these $\frac{1}{k}$
terms with the coefficients of the logarithmic divergences in the
corresponding diagrams.  As a result, by computing the divergences in
the bosonic and fermionic diagrams, we obtain a check on
eq.~(\ref{phasedeficit}), to leading order in $\frac{1}{k}$ for $k$ large.

Of course we must add back all that we have subtracted,
together with the contribution from the counterterm.  To do so we must
consider renormalization.  We will continue to use our simple 
renormalization scheme, in which we introduce only the subtraction
\begin{equation}
        {\cal L} \to {\cal L}  - C U^{\prime\prime}(\phi) U(\phi)
        - C U^{\prime\prime\prime}(\phi) \bar \Psi \Psi
        \label{counterterm1}
\end{equation}
which is equivalent to
\begin{equation}
        U(\phi) \to U(\phi) + \frac{\lambda}{m^2} C U^{\prime\prime}(\phi)
        \label{counterterm2}
\end{equation}
and thus preserves supersymmetry.  We fix the coefficient $C$ by
requiring that the boson tadpole (which includes contributions from both
boson and fermion loops as we have described above) vanish.  
In this scheme, the counterterm completely cancels the terms we have
subtracted from eq.~(\ref{unreg2}), so there is nothing to add back in.
In the sine-Gordon theory, this scheme also makes the physical mass of the
boson equal to $m$, while in the $\phi^4$ theory, there is a one-loop
correction to the physical mass of the boson from the diagram with two
three-boson vertices, giving a physical mass of
$m-\frac{\lambda}{4m\sqrt{3}}$ \cite{vN2}.  For us it is more important
to guarantee that the tadpole graphs vanish, assuring us that we have
chosen the correct vacuum for the theory, than to have the physical
mass equal to the Lagrangian parameter $m$; for the sine-Gordon case
we happen to be able to do both at once.  Furthermore, such
renormalization conditions can be applied uniformly to arbitrary $U(\phi)$.

Thus the effect of regularization and renormalization in 
our renormalization scheme is to subtract
\begin{equation}
        \delta^{(1)}(k) = \delta^{(1)}_B(k) - \delta^{(1)}_F(k) = 
        \frac{m}{k}
\end{equation}
from the difference of the boson and fermion phase shifts, giving
\begin{eqnarray}
\Delta H &=& \half\sum_j (\omega_j^B - m) - \half\sum_j 
(\omega_j^F - m) \cr
&& + \int_0^\infty \frac{dk}{2\pi} (\omega-m) \left(\frac{d
\delta_B}{dk} -  \frac{d \delta_F}{dk} - \frac{d\delta^{(1)}}{dk}\right) \cr
&=& -\frac{m}{4} + \int_0^{\infty} \frac{dk}{2\pi} (\omega-m) \frac{d}{dk} 
\left(\tan^{-1} \frac{m}{k} - \frac{m}{k}\right) =  -\frac{m}{2\pi}
\label{Hphase}
\end{eqnarray}
for both the kink and sine-Gordon soliton.  This result agrees with
\cite{vN2} and \cite{Sch}, and disagrees with \cite{many}, \cite{vN1},
\cite{KR}, and \cite{IM}. As pointed out in \cite{vN2}, in the case of the
sine-Gordon soliton,  it also agrees with the result obtained from the
Yang-Baxter equation assuming the factorization of the S-matrix \cite{YB}.

We note that in the end this result depended only on
eq.~(\ref{phasedeficit}) and its implications for Levinson's theorem.
Thus, since eq.~(\ref{phasedeficit}) holds in general for
antisymmetric soliton solutions, eq.~(\ref{Hphase}) gives the one-loop
correction to the energy in our renormalization scheme of any
supersymmetric soliton that is antisymmetric under reflection.

\section{Supersymmetry algebra and the central charge}

Our second application of the apparatus we have developed is to
compute the one-loop quantum correction to the central charge in the 
presence of the kink or sine-Gordon soliton.

First we summarize the supersymmetry algebra.  We define
\begin{equation}
  Q_\pm = \frac{(1\mp i\gamma^1)}{2} \frac{m^2}{\lambda} \int
  (\slash\hspace{-0.5em}\partial\phi + iU)\gamma^0\Psi \,dx
  = \frac{m^2}{\lambda} \int\left(\Pi \Psi_\pm + 
    (\phi^\prime \pm U)\Psi_\mp \right) \,dx
\end{equation}
where $\Psi_\pm = \frac{1\mp i\gamma^1}{2} \Psi$ and $Q_\pm = 
\frac{1\mp i\gamma^1}{2} Q$.  Using the canonical equal-time 
(anti)commutation relations, we have
\begin{eqnarray}
        \frac{m^2}{\lambda} \{ i\Psi_\pm(x),~\Psi_\pm(y) \} = i\delta(x-y) \cr
        \frac{m^2}{\lambda} [ \phi(x),~\Pi(y) ] = i\delta(x-y)
\end{eqnarray}
where $\Pi = \dot\phi$ is the momentum conjugate to $\phi$ and 
all other (anti)commutators vanish.  The supersymmetry algebra is
\begin{equation}
        \{ Q_\pm,~Q_\pm \} = 2 H \pm 2Z ~~~~~~~ \{ Q_+, Q_- \} = 2P,
\end{equation}
where $H$, $P$, and $Z$ are given classically by
\begin{eqnarray}   
H &=& \frac{m^2}{2\lambda} \int \left(\Pi^2 + (\phi^\prime)^2 + U^2
+ i (\Psi_-\Psi_+^\prime + \Psi_+ \Psi_-^\prime) + 
2i U^\prime \Psi_- \Psi_+ \right) \, dx \cr P &=& \frac{m^2}{\lambda} 
\int \left( \Pi\phi^\prime + \frac{i}{2} (\Psi_+ 
\Psi_+^\prime + \Psi_- \Psi_-^\prime) \right) \, dx \cr Z &=&
\frac{m^2}{\lambda} \int \phi^\prime U \, dx.
\label{classicalops}    
\end{eqnarray}
It is easy to check that $H$ is the same Hamiltonian as would be
determined canonically from eq.~(\ref{LaG}).

At the classical level, using eq.~(\ref{virial}),
\begin{equation}
        H_{\rm cl} = \frac{m^2}{2\lambda} \int \left(\phi_0^\prime(x)^2 
        + U(\phi_0)^2\right) dx
        = \mp \frac{m^2}{\lambda} \int U(\phi_0(x)) \phi_0^\prime(x)
        = \mp Z_{\rm cl},
\end{equation}
for the soliton and antisoliton respectively.

The hermiticity of $Q_\pm$ gives the BPS bound on the expectation
values of $H$ and $Z$ in any quantum state:
\begin{equation}
\langle H \rangle \geq \left| \langle Z \rangle \right|.
\label{BPSbound}
\end{equation}
Classically, the values of $H$ and $|Z|$ are equal so this bound is saturated.
We have found a negative correction to $H$ at one-loop, so if there is
no correction to $Z$, eq.~(\ref{BPSbound}) will be violated.

To unambiguously compute the corrections to the central 
charge for a soliton, it is easier to consider corrections to 
$Q_{+}^{2}=H+Z$, which is zero classically (for the antisoliton 
we should consider $Q_{-}^{2}=H+Z$).  One reason to 
consider $Q_{+}^{2}$ rather than $H$ and $Z$ separately is that
this quantity is finite and independent of the renormalization 
scheme.  Using eq.~(\ref{counterterm1}) and eq.~(\ref{counterterm2})
we see explicitly that the contribution from the counterterm cancels:
\begin{equation}
\Delta H_{\rm ct} = C \int  U^{\prime\prime}(\phi_0) U(\phi_0)
\, dx = - C \int U^{\prime\prime}(\phi_0) \phi_0^\prime \, dx =
 - \Delta Z_{\rm ct}
\label{countertermHZ}
\end{equation}
(and we only need consider the tree-level contribution since the counterterm
coefficient $C$ is already order $\lambda^0$).

Next we expand $\phi(x) = \phi_0(x) + \eta(x)$, where the soliton solution
$\phi_0$ is an ordinary real function of $x$.  Neglecting terms of
order $\eta^3$ and higher (which give higher-loop corrections), we obtain
\begin{eqnarray}
\langle H+Z\rangle_{\phi} &=& \frac{m^2}{2\lambda} \int \left\langle \Pi^2 +
\left[\left( \frac{d}{dx} + U^\prime(\phi_0)\right)\eta\right]^2
\right. \cr && \left.
+ i \Psi_+ \left( \frac{d}{dx} - U^\prime(\phi_0) \right) \Psi_- 
\right. \cr && \left.
+ i \Psi_- \left( \frac{d}{dx} + U^\prime(\phi_0) \right) \Psi_+
\right\rangle_{\phi} \,dx,
\end{eqnarray}
where $\langle\rangle_\phi$ denotes expectation value in 
the classical soliton background.

To evaluate this expression, we decompose the fields $\eta$ and $\Psi$
using creation and annihilation operators for the small oscillations
around $\phi_0$.  The small oscillation modes will be given in terms of the 
eigenmodes of the bosonic potentials $V_\ell(x)$ and
$\tilde V_\ell(x) = \frac{1}{\ell^2} V_{\ell-1}(\frac{x}{\ell})$.
For any mode $\eta_k(x)$ of $V_\ell(x)$ with nonzero energy
$\omega_k = \sqrt{k^2 + m^2}$, there is a mode $\tilde \eta_k(x)$
of $\tilde V_\ell(x)$ with the same energy, related by
\begin{eqnarray}
        \omega_k \tilde \eta_k(x)&=& i\left( \frac{d}{dx} + U^\prime(\phi_0)
        \right) \eta_k\nonumber\\
        \omega_k \eta_k(x) &=& i\left( \frac{d}{dx} - U^\prime(\phi_0)
        \right) \tilde\eta_k.
        \label{etatilde}
\end{eqnarray}

We use these wavefunctions to obtain
\begin{eqnarray}
\eta(x) &=& \sqrt{\frac{\lambda}{m^2}}
\int \frac{dk}{\sqrt{4\pi\omega_{k}} } \left(a_k \eta_k(x)
e^{-i\omega_k t} + a_k^\dagger \eta_k^\ast(x) e^{i\omega_k  t}\right) \cr
&& + \sqrt{\frac{\lambda}{m^2}} \eta_{\omega=0}(x) a_{\omega=0} \cr
\Psi(x) &=& \sqrt{\frac{\lambda}{m^2}} \int
\frac{dk}{\sqrt{4\pi\omega_{k}}} \left(b_k \psi_k(x) e^{-i\omega_k t} + 
b_k^\dagger \psi_k^\ast(x) e^{i\omega_k t}\right) \cr
&& + \sqrt{\frac{\lambda}{m^2}} \psi_{\omega=0}(0) b_{\omega=0}
\end{eqnarray}
where $\eta_{-k}(x) = \eta_k^\ast(x)$, the creation and annihilation
operators obey
\begin{equation}
[a_k,~a^{\dagger}_{k^\prime}] = \{b_k,~b^{\dagger}_{k^\prime}\} =
\delta(k-k^\prime)
\end{equation}
with all other (anti)commutators vanishing, and
\begin{equation}
        \psi_k(x) = \sqrt{\omega_k} \left( \matrix{ \eta_k(x) \cr 
        \tilde \eta_k(x) } \right).
\end{equation}
We note that the integral over $k$ also 
includes discrete contributions from the bound states (which 
correspond to imaginary values of $k$).  These are understood to give
discrete contributions to the results that follow (with Dirac 
delta functions replaced by Kronecker delta functions appropriately).
However, we have explicitly indicated the contribution from the bound
states at $\omega=0$ following \cite{JR}.  

We normalize the wavefunctions $\eta_k$ such that
\begin{equation}
\int \frac{dk}{2\pi} \eta_k(x)^\ast \eta_k(y) = \delta(x-y)
\end{equation}
which implies
\begin{equation}
\int \frac{dk}{2\pi} \tilde \eta_k(x)^\ast \tilde\eta_k(y) = \delta(x-y).
\end{equation}
With this normalization, the fields $\eta$ and $\Psi$ obey canonical
commutation relations.  Elementary algebra yields
\begin{eqnarray}
  i\left( \frac{d}{dx} + U^\prime(\phi_0)\right) \eta &=& 
  \sqrt{\frac{\lambda}{m^2}}
  \int \frac{dk\sqrt{\omega_k}}{\sqrt{4\pi}} \left(a_k \tilde \eta_k(x)
    e^{-i\omega_k t} - a_k^\dagger \tilde \eta_k(x)^\ast e^{i\omega_k t}\right) \cr
  i\Pi &=& \sqrt{\frac{\lambda}{m^2}}
  \int \frac{dk \sqrt{\omega_k}}{\sqrt{4\pi}} \left(a_k \eta_k(x)
    e^{-i\omega_k t} - a_k^\dagger \eta_k(x)^\ast e^{i\omega_k t}\right) \cr
  \Psi_+ &=& \sqrt{\frac{\lambda}{m^2}}
    \int \frac{dk}{\sqrt{4\pi}} \left(b_k \eta_k(x) e^{-i\omega_k t} +
      b_k^\dagger \eta_k(x)^\ast e^{i\omega_k t}\right) \cr
      && + \sqrt{\frac{\lambda}{m^2}} \eta_{\omega=0}(x) b_{\omega=0}\cr
  \Psi_- &=& \sqrt{\frac{\lambda}{m^2}} \int \frac{dk}{\sqrt{4\pi}} \left(b_k
    \tilde \eta_k(x)   e^{-i\omega_k t} + b_k^\dagger \tilde
    \eta_k(x)^\ast e^{i\omega_k     t}\right) \cr 
  i \left( \frac{d}{dx} + U^\prime(\phi_0) \right) \Psi_+ &=& 
  \sqrt{\frac{\lambda}{m^2}} \int \frac{\omega_k dk}{\sqrt{4\pi}}
   \left(b_k \tilde \eta_k(x)    e^{-i\omega_k t} -
  b_k^\dagger \tilde \eta_k(x)^\ast e^{i\omega_k t}\right) \cr
  i \left( \frac{d}{dx} - U^\prime(\phi_0) \right) \Psi_- &=&
 \sqrt{\frac{\lambda}{m^2}} \int \frac{ \omega_k dk}{\sqrt{4\pi}}
  \left(b_k \eta_k(x)  e^{-i\omega_k t} - b_k^\dagger \eta_k(x)^\ast
    e^{i\omega_k t}\right)
    \label{fieldexpansions}    
\end{eqnarray}
and we find
\begin{eqnarray}
        \langle H+Z \rangle_{\phi} &=&
        \int dx \int \frac{dk}{8\pi} \omega_k \left|\eta_k(x) \right|^2
        + \int dx \int \frac{dk}{8\pi} \omega_k \left| \tilde
        \eta_k(x) \right|^2 \cr
        && - \int dx \int \frac{dk}{8\pi} \omega_k \left| \tilde \eta_k(x)
         \right|^2 - \int dx \int \frac{dk}{8\pi} \omega_k \left|
         \eta_k(x) \right|^2  = 0
 \label{saturation}
\end{eqnarray}
and the BPS bound remains saturated.  
(If we instead considered an antisoliton, we would find the same
result for $\langle  Q_{-}^{2}\rangle_{\bar\phi} = \langle H - Z
\rangle_{\bar\phi}$, with the roles of $\Psi_+$ and $\Psi_-$
reversed.)  Our result disagrees with \cite{vN2} and \cite{IM}, which
claim that there is no correction to the central charge at one
loop in this renormalization scheme. We note that this result did not
depend on any specific properties of $U$, so it holds for any
supersymmetric soliton satisfying eq.~(\ref{virial}).

The second line of eq.~(\ref{saturation}) is simply the
unregulated fermionic contribution to the energy, and is explicitly
equal to minus the average of the contributions from the bosonic
potentials $V_\ell$ and $\tilde V_\ell$, in agreement with what we
found above.  As a final consistency check, we recalculate the full one-loop 
correction to the energy and central charge using our expansion in terms of
quantum fields.  For $\Delta H$ we obtain, again neglecting $\eta^3$ terms,
\begin{eqnarray}
 \Delta H &=& \langle H \rangle_\phi - H_{\rm cl} \cr  
  &=& \frac{m^2}{2\lambda} \int \left\langle \Pi^2 +
  \eta\left( -\frac{d^2}{dx^2} +  U^\prime(\phi_0)^2 +
  U(\phi_0)U^{\prime\prime}(\phi_0)\right)\eta \right.\cr
  && \left.
  + i \Psi_+ \left( \frac{d}{dx} - U^\prime(\phi_0) \right) \Psi_-
  +  i \Psi_- \left( \frac{d}{dx} + U^\prime(\phi_0) \right) \Psi_+
  \right\rangle_{\phi} \,dx \cr
   &=& \Delta H_{\rm ct} \cr
   && + \int dx \int \frac{dk}{4\pi} \omega_k \left| \eta_k(x) \right|^2 
   - \int dx \int \frac{dk}{8\pi} \omega_k \left( \left| \tilde \eta_k(x)
   \right|^2 +  \left|\eta_k(x) \right|^2  \right).
   \label{Hfield}
\end{eqnarray}
To relate this expression to our phase shift formalism, we consider
the Green's function for the bosonic field
\begin{eqnarray}
 G(x,y,t) &=& i{\rm T~} \langle \eta(x,t) \eta(y,0) \rangle \cr
 &=& i \int \frac{dk}{4\pi\omega_k} \left( 
 e^{i\omega_k t} \eta_k^\ast(x) \eta_k(y) \Theta(t)
 \right. \cr && \left.
\qquad\qquad + e^{-i\omega_k t} \eta_k(x) \eta_k^\ast(y) \Theta(-t) \right)
\end{eqnarray}
and its Fourier transform
\begin{equation}
        G(x,y,\omega) = \int G(x,y,t) e^{i\omega t} dt = \int 
        \frac{dk}{2\pi}\left(\frac {\eta_k(x) \eta_k^\ast(y)} {\omega^2 
        - \omega_k^2 - i\epsilon}\right)
\end{equation}
whose trace gives the density of states according to
\begin{equation}
        \rho_B(\omega) = {\rm Im } \:\frac{2\omega}{\pi} \int 
        G(x,x,\omega) \, dx
\end{equation}
giving as a result
\begin{equation}
    \rho_B(k) = \frac{1}{\pi} \int dx |\eta_k(x)|^2.
\end{equation}
Similarly for the fermions we find
\begin{equation}
        \rho_F(k) = \frac{1}{2\pi}\int dx \left(|\eta_k(x)|^2 + 
        |\tilde \eta_k(x)|^2\right).
\end{equation}
These results enable us to verify that eq.~(\ref{Hfield}) is in
agreement with eq.~(\ref{Hphase}).

In the exact same way, we can calculate the correction to $Z$ directly. We
start from the classical expression for $Z$ in eq.~(\ref{classicalops}) and
expand about the classical solution $\phi = \phi_0$, giving
\begin{eqnarray}
\Delta Z
&=& \langle Z \rangle_\phi - Z_{\rm cl} \cr  
&=& \Delta Z_{\rm ct} + \frac{m^2}{\lambda}\int
\left\langle U^\prime \eta \eta^\prime - \half UU^{\prime\prime}\eta^2
\right\rangle_{\phi} \, dx \cr
&=& \Delta Z_{\rm ct} + \frac{m^2}{2\lambda}
\int\left\langle\left((\frac{d}{dx} + U^\prime)\eta\right)^2 -
(\eta^\prime)^2 -
\eta^2(U^\prime)^2 - UU^{\prime\prime}\eta^2
\right\rangle_{\phi} dx.
\end{eqnarray}
After substituting the expansions of eq.~(\ref{fieldexpansions}) we obtain
\begin{eqnarray}
\Delta Z &=& \Delta Z_{\rm ct} + \int dx \int \frac{dk}{8\pi} 
\omega_k \left| \tilde\eta_k(x) \right|^2 - \int dx \int 
\frac{dk}{8\pi} \omega_k \left| \eta_k(x) \right|^2 \cr
&=& \frac{1}{4} \sum_j (\tilde \omega_j - m) - \frac{1}{4} \sum_j 
(\omega_j - m) \cr
&& + \int \frac{dk}{4\pi} (\omega_k - m) 
\frac{d}{dk} \left(\tilde\delta_l(k) - \delta_l(k) + 
2\delta^{(1)}(k) \right) \cr
 &=& \frac{m}{4} - \int 
\frac{dk}{2\pi} (\omega-m) \frac{d}{dk} \left(\tan^{-1} 
\frac{m}{k} - \frac{m}{k}\right) = \frac{m}{2\pi} = -\Delta H.
\end{eqnarray}

Another work \cite{SVV} has explained the difference between our
result for the central charge and the result obtained in Ref.~\cite{vN2} and
predecessors.  The earlier works found $\Delta Z = 0$ based on an
argument that involves direct manipulation of the $Z$ operator.
However, Ref.~\cite{SVV} showed that manipulations of this type are
only valid if the operator is augmented with an anomalous correction
of order $\hbar$.

We have systematically avoided the questionable manipulations that
would have led to $\Delta Z = 0$ by computing only matrix elements,
where we have the phase shift formalism available to  guide us.  First
we developed an unambiguous renormalization  procedure for $H$ based
on physical quantities.  Classical BPS saturation defines the operator
$Z$ at tree level.  Once we have fixed a renormalization scheme in our
computation of $H$, the expectation value of $Z$ (or any other
physical quantity) is determined.  We then carried out the one-loop
computation of $Z$ exactly in parallel to the computation of $H$ so
that no new ambiguities could arise, and found that BPS saturation is
maintained at one loop, as confirmed in Ref.~\cite{SVV}.  Our use of
Levinson's theorem and the Born approximation also prevented the
appearance of the spurious linear divergences found in Ref.~\cite{vN2}.

Thus we consistently included the effects of the SVV anomaly in both
$H$ and $Z$, though of course we found only a particular matrix
element instead of the full operator.  However, by working in the
continuum (without boundaries), we avoided the difficulties that
Ref.~\cite{SVV} faced in separating the unphysical contributions at
the boundaries from the physical effects localized at the soliton.

\clearpage
\newpage

\chapter{Scalars in three dimensions}

Having fully analyzed models in one dimension, we now turn to three
dimensions.  We will continue to consider models with spherical
symmetry, so that we can use a partial wave decomposition.  Thus, in
addition to an integral over $k$, we will also have a sum over $\ell$
to consider.  On the other hand, the subtleties of the symmetric
channel in one dimension will be absent.  Theories in three dimensions
will also have stronger divergences than we found in one dimension.
Thus we will have to subtract more Born terms and add back in more
corresponding diagrams.  However, since all the theories we will
consider are renormalizable, only a finite number of Born subtractions
will be required.  We will again begin with a scalar model, but our
techniques will easily generalize to include fermions.

\section{Formalism}

We consider a renormalizable field theory with a real scalar field $\phi$
coupled to a charged scalar $\Psi$. We take the classical potential 
$V(\phi)\propto (\phi^2-v^2)^2$, and
$\Psi$ acquires a mass through spontaneous symmetry breaking. 
At the quantum level we put aside the
$\phi$ self-couplings and consider only the effects of the $\phi-\Psi$
interactions.  We further restrict ourselves to ${\cal O}(\hbar)$
effects in the quantum theory, which correspond to one-loop diagrams.

Our model is defined by the classical action
\begin{eqnarray}
S[\phi,\Psi]&=&\int d^4x\left\{ \half
(\partial_\mu\phi)^2-\frac{\lambda}{4!}(\phi^2-v^2)^2 +
\partial_\mu\Psi^\ast\partial^\mu\Psi - g\Psi^\ast\phi^2\Psi
\right. \cr \nonumber  && \left. 
+ a (\partial_\mu\phi)^2 - b (\phi^2-v^2) - c (\phi^2-v^2)^2 
\phantom{\half}\right\}\ ,
\label{I.1}
\end{eqnarray}
where we have separated out the three counterterms necessary for
renormalization and written them in a convenient form.
At one-loop order in $\Psi$, these are the only
counterterms required.

We quantize around the classical vacuum $\phi=v$ and define
$h=\phi-v$, so that
\begin{eqnarray}
\label{I.1a}
S[h,\Psi] &=& \int d^4 x \left\{\half(\partial_\mu h)^2 - \frac{m^2}{8v^2}
(h^2 + 2vh)^2  
- g(h^2+2vh)\Psi^\ast \Psi
\right. \cr \nonumber  && \left. 
+ \partial_\mu\psi^\ast\partial^\mu\Psi -  M^2\psi^\ast\Psi
\right. \cr \nonumber  && \left. 
+ a(\partial_\mu h)^2 - b(h^2+2hv) - c(h^2+2hv)^2 
\phantom{\half}\right\}
\end{eqnarray}
where $M=\sqrt{g}v$ is the $\Psi$ mass and $m^2 = \lambda v^2/3$
is the $h$ mass.

The one-loop quantum effective action for $h$ is obtained by 
integrating out $\Psi$ to leading order in $\hbar$.
We are interested in time-independent field configurations
$h=h(\vec x)$, for which the effective action yields an
effective energy ${\cal E}[h]$ that has three parts:
\begin{equation}
  {\cal E}[h] = {\cal E}_{\rm cl}[h] + {\cal E}_{\rm ct}[h] 
   + {\cal E}_{\Psi}[h]\ ,
\end{equation}
where ${\cal E}_{\rm cl}[h]$ is the classical energy of $h$,
\begin{equation}
  {\cal E}_{\rm cl}[h] = \int d^3x\ \left\{ \half |\vec\nabla h|^2 +
  \frac{m^2}{8v^2} (h^2 + 2vh)^2 \right\}\ ,
\label{I.1b}
\end{equation} 
${\cal E}_{\rm ct}[h]$ is the counterterm contribution,
\begin{equation}
  {\cal E}_{\rm ct}[h] = \int d^3x\left\{ a |\vec\nabla h|^2 +
  b(h^2+2hv) + c(h^2+2hv)^2 \right\}\ ,
\label{I.1b1}
\end{equation}
and ${\cal E}_{\Psi}[h]$ is the one-loop quantum contribution from
$\Psi$.  ${\cal E}_{\rm ct}[h]$ and ${\cal E}_{\psi}[h]$
are divergent, but we will see explicitly that these divergences
cancel for any configuration $h(\vec x)$.

\begin{figure}
$$
\BoxedEPSF{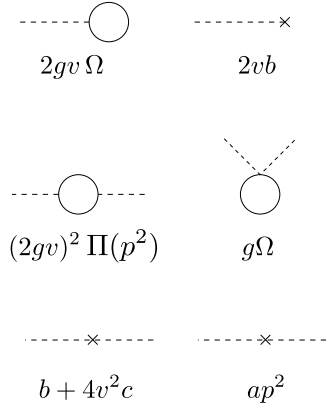 scaled 500}  
$$
\caption{One-loop diagrams.}
\label{figure31}
\end{figure}

We fix the counterterms by applying renormalization conditions
in the perturbative sector of the theory.  Having done so, we have
defined the theory for all $h(\vec x)$.  We choose the on-shell
renormalization conditions
\begin{equation}
  \Sigma_1=0,\quad \Sigma_2(m^2) = 0,\quad {\rm and}\quad
   \left.\frac{d\Sigma_2}{dp^2}\right|_{m^2} = 0,
\label{I.2a}
\end{equation}
where $\Sigma_1$ and $\Sigma_2(p^2)$ are the one- and two-point functions
arising only from the loop and counterterms as seen in
Fig.~\ref{figure31}. 
We denote the one-loop diagrams with one insertion by
$\Omega$ and with two insertions by $\Pi(p^2)$, and find
\begin{eqnarray}
  \Sigma_1 &= &2vg\Omega + 2vb\ ,\nonumber\\
  \Sigma_2(p^2) & = &(2vg)^2 \Pi(p^2) +g\Omega + b + (2v)^2c + a p^2\ .
\label{I.2b}
\end{eqnarray}
Defining
\begin{equation}
\Pi'(p^2) \equiv\frac{d\Pi(p^2)}{dp^2}\ ,
\end{equation}
the renormalization conditions eq.(\ref{I.2a}) then yield
\begin{equation}
a = -(2vg)^2 \Pi'(m^2)\ ,\ b = -g\Omega\ ,\ c = g^2(m^2 \Pi'(m^2)- 
\Pi(m^2))\ ,
\end{equation}
which we then substitute into the counterterm energy, eq.~(\ref{I.1b1}).

Now we consider the calculation of ${\cal E}_{\Psi}[ h]$. This energy
is the sum over zero point energies, $\half\hbar\omega$, of the modes of
$\Psi$ in the presence of $ h(\vec x)$,
\begin{equation}
  {\cal E}_\psi[ h] = \sum_\alpha \omega_k[ h]
\label{I.3}
\end{equation}
where $\omega_k$ are the positive square roots of the eigenvalues of
the small oscillations Hamiltonian,
\begin{equation}
\left(-\vec\nabla^2 + M^2 + g(h^2 + 2vh)\right)\psi_k = \omega_\alpha^2 \psi_k
\label{I.4}
\end{equation}
The fact that $\Psi$ is complex accounts for the absence of $\half$
in eq.~(\ref{I.3}).

${\cal E}_\Psi$ is highly divergent.  However our model is renormalizable and
therefore the counter\-terms fixed in the presence of the trivial $ h$
{\it must\/} cancel all divergences in ${\cal E}_\Psi$.  Rather than
attempt to regulate the divergent sum in eq.~(\ref{I.3}) directly, we
study the density of states that defines the sum. We can  isolate the
terms that lead to  divergences in ${\cal E}_\Psi$ and renormalize
them using conventional methods.  Thus our task is to generalize the
construction of eq.~(\ref{unrenorm}) to this case.

For fixed $h(\vec x)$ the spectrum of
$\hat H$ given in eq.~(\ref{I.4}) consists of a finite 
number (possibly zero) of normalizable bound states and a continuum
beginning at $M^2$, parameterized by 
$k$, with $E(k)=+\sqrt{k^2+M^2}$. Furthermore, $\hat H$ depends on $h$ only
through the combination
\begin{equation}
\chi = h^2+2hv\ ,
\end{equation}
so we can consider ${\cal E}_\Psi$ to be a functional
of $\chi$.  We restrict ourselves to spherically symmetric $h$.  Then
\begin{equation}
  {\cal E}_\Psi[\chi] = \sum_j E_j + \sum_\ell (2\ell +1) 
\int dk \rho_\ell(k) E(k)
\end{equation}
where $\rho_\ell(k)$ is the density of states in $k$ in the $\ell^{\rm th}$
partial wave and the $E_j$ are the bound state energies. $\rho_\ell(k)$ 
is finite, but the sum over $\ell$ and the integral over $k$ are 
divergent. Furthermore 
\begin{equation}
  \rho_\ell(k) = \rho_\ell^{\rm free}(k)
  + \frac{1}{\pi}\frac{d\delta_\ell(k)}{dk}\ ,
    \label{I.6}
\end{equation}
where $\delta_\ell(k)$ is the usual scattering phase shift for the
$\ell^{\rm th}$ partial wave, and $\rho_\ell^{\rm free}(k)$
is the free ($g=0$) density of states. This
relationship between the density of states and the derivative of the
phase shift is shown for example in \cite{Sch}.

At the outset, we subtract $\rho^{\rm free}(k)$ from the density 
of states since we wish to compare ${\cal E}_\Psi[\chi]$ to ${\cal E}_\Psi[0]$.
Viewing ${\cal E}_{\Psi}[\chi]$ as the sum of one loop
diagrams, we see that only the diagrams with one or two insertions 
of $g\chi$ are divergent. A diagram with $n$ insertions corresponds to
the $n^{\rm th}$  term in the Born expansion, so all possible
divergences can be eliminated by subtracting the first and second Born
approximations from the phase shifts that determine the
density of states. Standard methods allow us to construct the 
Born approximation for the phase shifts \cite{Schiff}, which is a
power series in the ``potential'' $g\chi$.

We define the combination
\begin{equation}
  \bar\delta_\ell(k) \equiv \delta_\ell(k) - \delta^{(1)}_\ell(k) -
  \delta^{(2)}_\ell(k)\ ,
   \label{I.7}
\end{equation}
where $\delta^{(1)}_\ell(k)$ and $\delta^{(2)}_\ell(k)$ are the first
and second Born approximations to $\delta_l(k)$. We then have
\begin{eqnarray}
  {\cal E}_\Psi [\chi]&=& \sum_j E_j + \sum_\ell (2\ell+1) 
\int_0^\infty dk
  \frac{1}{\pi} \frac{d\bar\delta(k)}{dk}E(k) + g\Omega 
  \int \frac{d^3\! p}{(2\pi)^3} \tilde \chi(0) \\ \nonumber
  &&\ + g^2 \int \frac{d^3\! p}{(2\pi)^3} \Pi(-\vec p\;\mbox{}^2)
  |\tilde\chi(\vec p)|^2
  \label{I.7a}
\end{eqnarray}
where 
\begin{equation}
\tilde \chi(\vec p) =\int d^3x \chi(\vec x) e^{-i\vec p \cdot \vec x}\ ,
\end{equation}
and likewise for $\tilde h(\vec p)$. Both  $\tilde h$ and 
$\tilde \chi$ are real and
depend only on $q\equiv|\vec p|$ for spherically symmetric $h$.
We have subtracted out the order $g$ and $g^2$ contributions by using
$\bar \delta_\ell(k)$ instead of $\delta_\ell(k)$, and added them back
in by using their explicit diagrammatic representation in terms of the
divergent constant $\Omega$ and the divergent function $\Pi(p^2)$.

We can now combine ${\cal E}_\Psi$ and ${\cal E_{\rm ct}}$ and obtain a
finite result:
\begin{eqnarray}
  {\cal E}_\Psi + {\cal E_{\rm ct}} &=& \sum_j E_j + \sum_\ell (2\ell+1)
  \int_0^\infty dk \frac{1}{\pi} \frac{d\bar\delta_{\ell}(k)}{dk}E(k)
  +\Gamma_2[h]
   \label{I.8}
\end{eqnarray}
where 
\begin{eqnarray}
  \Gamma_2[h] &=&
  \frac{g^2}{2} \int_0^\infty \frac{q^2 dq}{2\pi^2} \left[\left(\Pi(-q^2) 
  - \Pi(m^2) + m^2 \Pi'(m^2)\right) \tilde \chi(q)^2 \right] \cr
&& + \frac{g^2}{2} \int_0^\infty \frac{q^2 dq}{2\pi^2} \left[ 4v^2q^2
  \Pi'(-q^2) \tilde h(q)^2 \right] .
   \label{I.8a}
\end{eqnarray}
$\Pi$ is log divergent, but both 
$\{\Pi(-q^2) - \Pi(m^2)\}$ and $\Pi'$ are finite, so $\Gamma_2[h]$ is
finite as well.

Each term in the Born approximation to the phase shift goes to zero at
$k=0$, so by Levinson's theorem $\bar\delta_\ell(0) = \delta_\ell(0) =
\pi n_\ell$ where $n_\ell$ is the number of bound states with 
angular momentum $\ell$.  As $k\to\infty$, $\delta_\ell(k)$ falls off
like $\frac{1}{k}$, $\delta^{(1)}_\ell(k)$ falls off
like $\frac{1}{k}$, and $\delta^{(2)}_\ell(k)$ falls off
like $\frac{1}{k^2}$. Since the Born approximation becomes exact
at large $k$, $\bar\delta_\ell(k)$ falls like $\frac{1}{k^3}$. 
Thus we see that the first subtraction renders each 
integral over $k$ convergent.  The second subtraction makes the
$\ell$-sum convergent. We are then free to integrate by parts in
(\ref{I.8}), obtaining
\begin{equation} 
   {\cal E}[ h]= {\cal E}_{\rm cl}[h]
   +\Gamma_2[ h]
   - \frac{1}{\pi}\sum_\ell (2\ell+1)\int_0^\infty dk\:
   \bar\delta_\ell(k)\frac{k}{E(k)} +\sum_j (E_j-M) \ .
   \label{I.9}
\end{equation}
In this expression 
we see that each bound state contributes its binding energy, $E_j-M$,
so that the energy varies smoothly as we strengthen $h$ and bind more states.

\section{Calculational methods}

In this Section we describe the method that allows us to construct
${\cal E}[ h]$ as a functional of $ h$ and search for stationary
points. We now consider the calculation of each of the terms in 
eq.~(\ref{I.9})
in turn.  The classical contribution to the action is evaluated directly by
substitution into eq.~(\ref{I.1b}).
$\Gamma_2[h]$ of eq.~(\ref{I.8a}) is obtained from a Feynman diagram 
calculation,
\begin{eqnarray}
\Gamma_2 [h] &=& \frac{g^2}{(4\pi)^2}\int_0^\infty \frac{q^2dq}{(2\pi)^2}
\int_0^1 \left[
\log\frac{M^2+q^2x(1-x)}{M^2-m^2x(1-x)}
\right. \cr && \left. \qquad
-\frac{m^2 x(1-x)}{M^2-m^2x(1-x)} \right] \tilde \chi(q)^2 dx \cr
&& - \frac{g^2}{(4\pi)^2}\int_0^\infty \frac{q^2dq}{(2\pi)^2}
\int_0^1 \left[ \frac{x(1-x)}{M^2-m^2x(1-x)}
4v^2q^2\tilde{h}(q)^2 \right] dx .
\end{eqnarray}

The partial wave phase shifts and Born approximations are calculated
as follows.  The radial wave equation is
\begin{equation}
   -u_\ell^{\prime\prime} +\left[\frac{\ell(\ell+1)}{r^2}
   +g\chi(r)\right]u_\ell=k^2 u_\ell, \
\label{II.2}
\end{equation}
where $k^2>0$, and $\chi(r)\to 0$ as $r\to \infty$.
We introduce two linearly independent solutions to
eq.~(\ref{II.2}), $u_\ell^{(1)}(r)$ and $u_\ell^{(2)}(r)$, defined by
\begin{eqnarray}
   u_\ell^{(1)}(r) &=& e^{2i\beta_\ell(k,r)}r h_\ell^{(1)}(kr) \\ 
\nonumber
   u_\ell^{(2)}(r) &=& e^{-2i\beta_\ell^\ast(k,r)}r h_\ell^{(2)}(kr) 
\label{II.3}
\end{eqnarray}
where $h_\ell^{(1)}$ is the spherical
H\"ankel function asymptotic to $e^{ikr}/r$ as $r\rightarrow\infty$,
$h_\ell^{(2)}(kr) = h_\ell^{(1)\ast}(kr)$, and  
$\beta_\ell(k,r) \to 0$ as $r \to \infty$, so that 
$u_\ell^{(1)}(r)\to e^{ikr}$ and
$u_\ell^{(2)}(r)\to e^{-ikr}$ as $r\to\infty$.  The scattering
solution is then
\begin{equation}
u_{\ell}(r) = u_\ell^{(2)}(r) + e^{2i\delta_\ell(k)} u_\ell^{(1)}(r)\ ,
\end{equation}
and obeys $u_{\ell}(0) = 0$.  Thus we obtain
\begin{equation}
   \delta_\ell(k)=-2\:{\rm Re}\:\beta_\ell(k,0).
\label{II.4}
\end{equation}
Furthermore, $\beta_\ell$ obeys a
simple, non-linear differential equation obtained by substituting
$u^{(1)}_\ell$ into eq.~(\ref{II.2}),
\begin{equation}
   -i\beta_\ell^{\prime \prime}
   -2ikp_\ell(kr)\beta_\ell^\prime
   +2(\beta_\ell^\prime)^2+\half g\chi(r) = 0\ ,
\label{II.5}
\end{equation}
where primes denote differentiation with respect to $r$, and
\begin{equation}
        p_\ell(x)=\frac{d}{dx} \ln \left[ xh_\ell^{(1)}(x)\right]
\label{II.6}
\end{equation}
 is a simple rational function of $x$.

We solve eq.~(\ref{II.5}) numerically, integrating from $r=\infty$ 
to $r=0$
with $\beta_\ell^\prime(k,\infty)=\beta_\ell(k,\infty)=0$, to get the 
exact phase
shifts.  To get the Born approximation to $\beta_\ell$, we solve the 
equation
iteratively, writing $\beta_\ell = g\beta_{\ell 1} + g^2\beta_{\ell 2} +
\dots$, where $\beta_{\ell 1}$ satisfies
\begin{equation}
   -i\beta_{\ell 1}^{\prime \prime}
   -2ikp_\ell(kr)\beta_{\ell 1}^\prime+\half \chi(r) = 0
\label{II.7}
\end{equation}
and $\beta_{\ell 2}$ satisfies
\begin{equation}
   -i\beta_{\ell 2}^{\prime \prime} 
   -2ikp_\ell(kr)\beta_{\ell 2}^\prime 
   +2(\beta_{\ell 1}^\prime)^2 = 0.
\label{II.8}
\end{equation}
We can solve efficiently for $\beta_{\ell 1}$ and $\beta_{\ell 2}$
simultaneously
by combining these two equations into a coupled differential equation 
for the
vector $(\beta_{\ell 1},~\beta_{\ell 2})$.  This method is much faster than 
calculating the Born terms directly as iterated integrals in $r$ and will
generalize easily to a theory requiring higher-order counterterms.

Having found the phase shifts, we then use Levinson's theorem to
count bound states.  We then find
the energies of these bound states by using a shooting method to
solve the corresponding eigenvalue equation.  We use the effective
range approximation \cite{Schiff} to calculate the phase shift and
bound state energy near the threshold for forming an s-wave bound state.

\section{Results}

\begin{figure}
$$
\BoxedEPSF{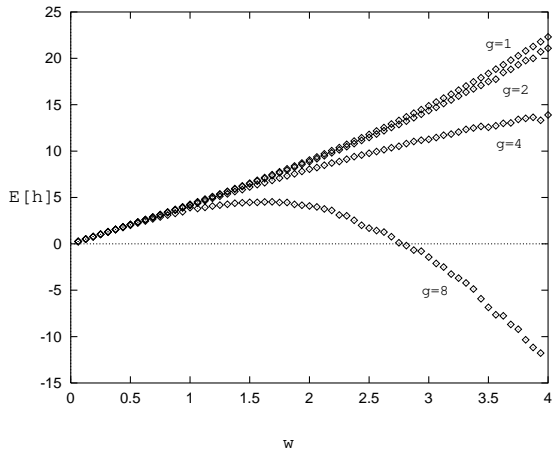 scaled 400}  
$$
\caption{${\cal E}[h]$ in units of $v$, for $d=1$ and 
$g=1,2,4,8$, as a function of $w$.}
\label{figure32}
\end{figure}

For the model at hand, we calculated the energy ${\cal E}[ h]$ for a 
two-parameter ($d$ and $w$) family of Gaussian backgrounds
\begin{equation}
h(r)=-dve^{-r^2 v^2/2w^2}\ .
\label{IV.1}
\end{equation}
In Fig.~\ref{figure32}, we show results which are representative of our
findings in general. We plot the energy of configurations with
fixed $d=1$ as a function of $w$, for $g=1,2,4,8$ (from top to bottom).
We note that to this order, for $g=8$ the vacuum is unstable to the 
formation of large $\phi=0$ regions.

To explore whether the charged scalar forms a non-topological soliton
in a given $\phi$ background, we add to ${\cal E}[ h]$ 
the energy of a ``valence'' $\Psi$ particle in the lowest bound state.
We then compare this total energy to $M$, the energy of the $\Psi$ particle
in a flat background, to see if the soliton is
favored.  This is the scalar model analog of $t$-quark bag
formation.

For fixed $g$ and $m$, we varied $h$ looking for
bound states with energy $E$ such that $E+{\cal E}[ h]<M$. However,
for those values of $g$ and $m$ where we did find such solutions, we
always found that by increasing $w$, we could make ${\cal E}[ h]<0$,
so that the vacuum is unstable, as we pointed out above in the case 
of $g=8$ in Fig.~\ref{figure32}.  Thus we find that if
we stay in the $g,m$ parameter region where the vacuum is stable, the
minimum is at $h=0$, so there are no nontopological solitons. 

Although we did not find a non-trivial solution at one-loop order
in this simple model, our calculation demonstrates the practicality
of our method in three dimensions.  We can effectively characterize
and search the space of field configurations, $h(r)$, while holding
the renormalized parameters of the theory fixed.  The same methods can be used
to study solitons in theories with richer structure in three dimensions.
                 
\section{Derivative expansion}

Our results are exact to one-loop order.  The derivative expansion, which is
often applied to problems of this sort, should be accurate for slowly varying
$h(r)$.  We found it useful to compare our results with the derivative 
expansion
for two reasons:  first, we can determine the range of validity (in $d$ 
and $w$)
of the derivative expansion; and second, where the derivative expansion is
expected to be valid, it provides a check on the accuracy of our 
numerical work
and C++ programming.  Where expected, the two calculations did agree to the
precision we specified (1 \%).

In our model, the first two terms in the derivative expansion of
the one-loop effective Lagrangian can be calculated to be
\begin{equation}\label{V.1}
{\cal L}_{\rm 1} = {\cal L}_{\rm ct} + \alpha z+\beta z^2
+ {g^2v^4\over 32\pi^2}\left[
(1+z)^2\ln (1+z)-z-{3\over2}z^2\right]+
{g\over48\pi^2v^2}{1\over 1+z} (\partial_\mu z)^2  \ ,
\end{equation}
where $z=g\chi/M^2=(h^2+2hv)/v^2$, $\alpha$ and $\beta$ are
cutoff-dependent constants, and ${\cal L}_{\rm ct}$ is the same
counterterm Lagrangian as we used in Sec. 2.  For $\phi^4$
scalar field theory a similar result was first derived in
\cite{Cole}. The last term above is proportional to $(\partial
h)^2$, and is completely cancelled by a finite counterterm that
implements the renormalization prescription of Sec. 2. In this
prescription, counterterms also cancel the $\alpha z$ and $\beta z^2$
terms above. Thus the ${\cal O}(p^2)$ derivative expansion for the  effective
energy, to be compared with the phase shift expression for ${\cal E}[h]$, is

\begin{eqnarray}
{\cal E}_{\rm DE}[h] &=& \int d^3x \left\{
\frac{(\vec{\nabla}h)^2}{2} +\frac{m^2}{8v^2}(h^2+2hv)^2 
\right. \cr && \left.
+\frac{g^2v^4}{32\pi^2}\left[ (1+z)^2\ln (1+z )-
z-{3\over2}z^2\right]\right\}\ .
\end{eqnarray}

\begin{figure}
$$
\BoxedEPSF{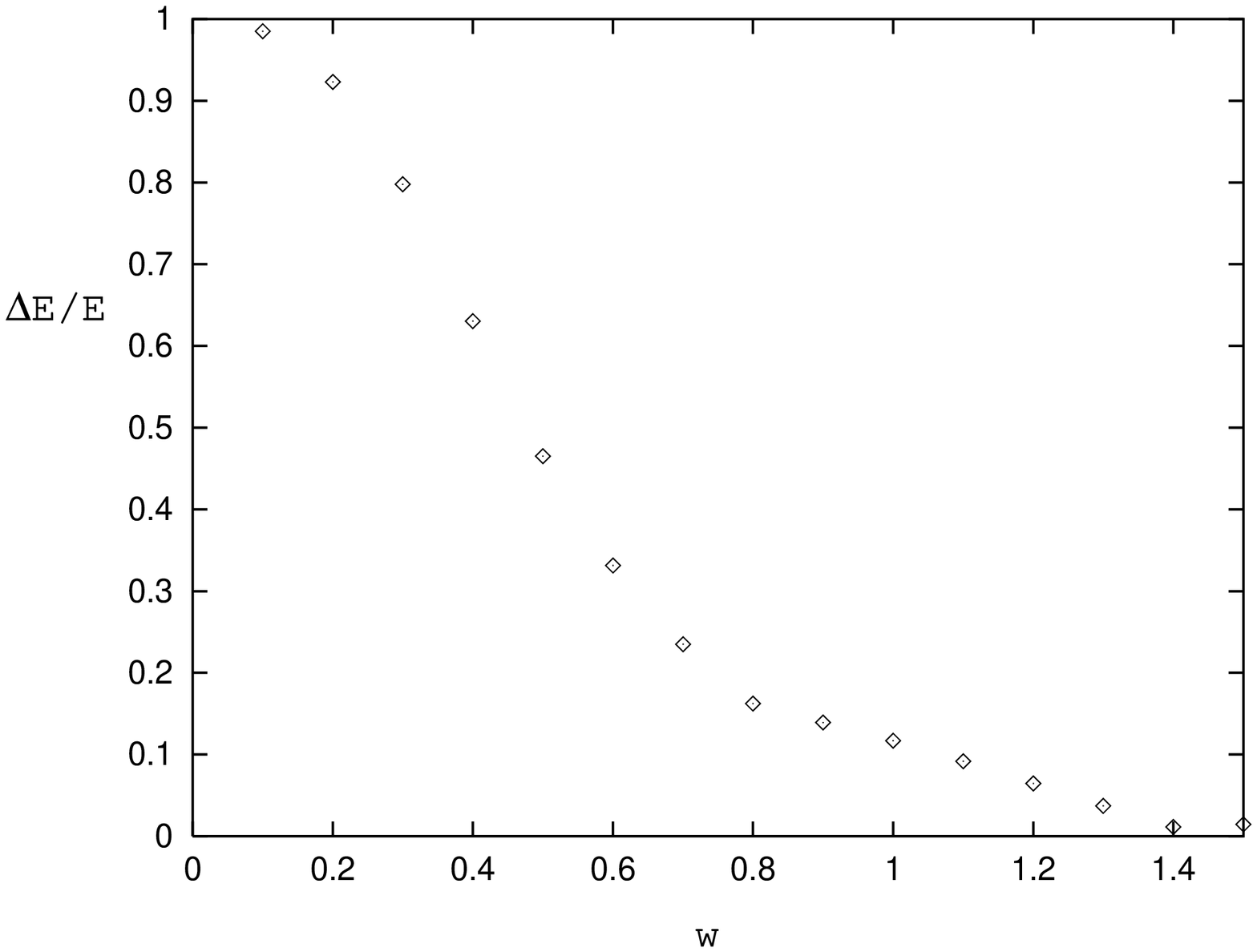 scaled 400}  
$$
\caption{$\left|( {\cal E} -{\cal E}_{\rm DE})/
({\cal E} - {\cal E_{ \rm cl}})\right|$
for $d=0.25$, $g=4$, as a function of $w$.}
\label{figure3}
\end{figure}

The results of the comparison with the phase shift method can 
be seen in Fig.~\ref{figure3} for $d = 0.25$,
and $g=4$. A similar pattern holds in general for other values of both
$d$ and $g$. As the width becomes larger, the two results merge. This is
as expected, since it is for large widths, and thus small gradients,
that we expect the derivative expansion to yield accurate results. As the
width tends to zero, both results tend to zero, and the fact that the plot 
tends to 1 simply indicates that the derivative expansion result goes to
zero faster than the phase shift result.

\clearpage
\newpage

\chapter{Fermions in three dimensions}

It is now an easy matter to extend these three dimensional results to
fermions.  We continue to consider a classical background $\phi$, now
coupled to a Dirac fermion $\Psi$.  Our Lagrangian becomes
\begin{eqnarray}
{\cal L} &=& \half
(\partial_\mu\phi)^2-\frac{\lambda}{4!}(\phi^2-v^2)^2 +
i \bar \psi \partial \hspace{-0.5em} \slash \psi - g\phi\bar\psi\psi
\cr && 
 + a (\partial_\mu\phi)^2 - b (\phi^2-v^2) - c (\phi^2-v^2)^2 
\phantom{\half}.
\end{eqnarray}
The small oscillations are now given by the Dirac equation,
\begin{equation}
\gamma^0 \left(-i\gamma^i \partial_i + m + gh\right) \psi_k = \omega \psi_k
\label{Dirac3}
\end{equation}
where $h = \phi-v$ and $m=gv$.  We will continue to work with a symmetric
background $h$, so we can decompose our small oscillations into
channels according to angular momentum.

Letting $\alpha^i = \gamma^0\gamma^i$ and $\beta = \gamma^0$, we will
work in the basis
\begin{equation}
\alpha^k = \left(\matrix { 0 & \sigma_k \cr \sigma_k & 0 } \right)
\hbox{~~~~and~~~~}
\beta = \left(\matrix { 1 & 0 \cr 0 & -1 } \right)
\end{equation}
where the $\sigma_k$ are the standard Pauli matrices.  We define the operator
\begin{equation}
Q = \beta(\vec \sigma \cdot \vec L)
\end{equation}
where $\vec L = \vec r \times \vec p$.  This operator commutes with
the Hamiltonian, and is related to the total angular momentum operator
$\vec J = \vec L + \half \vec \sigma$ by $Q^2 = \vec J^2 + \frac{1}{4}$.
We can thus restrict to states with a particular eigenvalue $q$ of the
operator $Q$, with $q=\pm 1,\pm 2,\pm 3\dots $.  We also introduce
the radial momentum operator
\begin{equation}
p_r = -i\left( \frac{d}{dr} + \frac{1}{r}  \right)
\end{equation}
which allows us to rewrite eq.~(\ref{Dirac3}) as
\begin{equation}
\left( \alpha_r p_r + \frac{i}{r} \alpha_r\beta q +
\beta(m+gh(r))\right) \psi_k = \omega \psi_k
\end{equation}
where $\alpha_r = \hat r \cdot \vec \alpha$.
The solutions to this equation will be of the form
\begin{equation}
\psi^+_k = \frac{1}{r} \left( \matrix{F(r) {\cal Y}^M_{(q-1) j} \cr
	iG(r) {\cal Y}^M_{q j} } \right)
\label{Diracwf1}
\end{equation}
for $q>0$ and
\begin{equation}
\psi^-_k = \frac{1}{r} \left( \matrix{F(r) {\cal Y}^M_{|q| j} \cr
	-iG(r) {\cal Y}^M_{(|q|-1) j} } \right)
\label{Diracwf2}
\end{equation}
for $q<0$, where ${\cal Y}^M_{\ell j}$ is the two-component spinor spherical
harmonic corresponding to a state with total angular momentum $j$,
orbital angular momentum $\ell$, and a $z$ component of angular momentum $M$.
These two solutions are eigenstates of parity with eigenvalues $\pm
(-1)^q$ respectively.  The functions $F$ and $G$ obey the coupled
radial equations
\begin{eqnarray}
\left((\omega - m) - gh(r)\right) F + \frac{dG}{dr} + \frac{q}{r}G &=& 0 \cr
\left((\omega + m) + gh(r)\right) G - \frac{dF}{dr} + \frac{q}{r}F &=& 0.
\end{eqnarray}
If we set $h(r) =0$, $F$ and $G$ obey the second-order equations
\begin{eqnarray}
\left(-\frac{d^2}{dr^2} + \frac{q(q-1)}{r^2}\right) F &=& k^2 F \cr
\left(-\frac{d^2}{dr^2} + \frac{q(q+1)}{r^2}\right) G &=& k^2 G
\end{eqnarray}
where $\omega = \sqrt{k^2 + m^2}$.  Thus the free solutions for $F$
and $G$ will be spherical harmonics $j_\ell(kr)$, with $\ell$ equal to
the orbital angular momentum of the spherical harmonic that the radial
function multiplies in eq.~(\ref{Diracwf1}) or eq.~(\ref{Diracwf2}).

In each channel, it is convenient to work with the full
second-order equation that we obtain by squaring the Dirac equation.
Any solution to the Dirac equation must solve this equation as well,
and if we choose our boundary conditions so that the Dirac equation is
solved, a solution to the second-order equation will automatically
solve the Dirac equation.  To facilitate our numerical computation, we
will focus on the upper component, since its equation will involve the
quantity $\omega+m$ rather than $\omega - m$, and the former is easier
to deal with numerically at small $k$.  (The small $k$ limit is also
the nonrelativistic limit, and in this limit the upper component is
large compared to the lower component, which makes it easier to deal
with numerically; if we were investigating negative energy solutions
it would be easier to use the lower component.)  We will therefore
label our channels by parity (as above) and by  $\ell = 0,1,2,\dots$,
the orbital angular momentum of the upper component.  The 
total spin $j$ is then equal to $\ell+\half$ for positive parity and
$\ell-\half$ for negative parity.  (Only positive parity exists for
$\ell=0$.)

Plugging $F(r) = e^{2i\beta(r)} r h^{(1)}_\ell(kr)$ into the full
second-order equation for $F$ gives
\begin{equation}
-i{\beta^+_\ell}^{\prime \prime} - 2ikp_\ell(kr){\beta^+_\ell}^\prime
+2({\beta^+_\ell}^\prime)^2 + \half \left(g^2\chi + g\frac{
h^\prime(r)(k p_\ell(kr)- \frac{\ell+1}{r} +
i{\beta^+_\ell}^\prime)}{(E+m+gh(r))} \right) = 0
\end{equation}
in the positive parity channel, and
\begin{equation}
-i{\beta^-_\ell}{}^{\prime \prime} - 2ikp_\ell(kr) {\beta^-_\ell}^\prime
+2({\beta^-_\ell}^\prime)^2 + \half \left(g^2\chi + g\frac{
h(r)^\prime(k p_\ell(kr) + \frac{\ell}{r} +
{i\beta^-_\ell}^\prime)}{(E+m+gh(r))} \right) = 0 
\end{equation}
in the negative parity channel.  As in the bosonic case, we have defined
\begin{equation}
\chi = h^2+2hv.
\end{equation}
Again, the phase shift is given by
\begin{equation}
\delta^\pm_\ell(k) = -2\:{\rm Re}\:\beta^\pm_\ell(k,0).
\end{equation}
We also have stronger divergences than in the bosonic case, so we must
subtract the first four terms in the Born approximation and add them
back in as diagrams.  As a result, the numerical computation is
somewhat more complicated; nonetheless, the basic method is the same:
we iterate the differential equations for the phase shifts to obtain
the Born approximations, which again cancel the divergences at large
$k$ and $\ell$ (though for small $k$ and $\ell$, the Born
approximation may deviate widely from the exact phase shift).
As in the boson model, the Born approximations to the phase shift go
to zero at $k=0$, while the exact phase shifts go to $\pi$ times the
number of bound states in that channel.  And, as we would expect,
there is an overall Fermi minus sign in front of the entire result.

This computation adds the full Casimir energy to the computation of
\cite{Selipsky}; as indicated above, to consistently compute the
effect of the shift in the valence quark level requires that we
include the Casimir energy as well.  However, in this model, although
the quantitative results are modified, the qualitative conclusions for
moderate values of $g$ are unchanged.

An interesting generalization of this model is to add isospin, making the
background field into a complex 2-component vector.  We can then
construct ``hedgehog'' configurations that have nontrivial topology.
These configurations are no longer $CP$ invariant.  In particular,
they can have spectral asymmetry, which is reflected by a fermion
level crossing during the process of adiabatically constructing the
configuration from the trivial vacuum.  If a level crossing does
occur, the configuration carries fermion number and we do not need to
fill an energy level at all.  Even if a level crossing does not
occur, the level that is heading toward zero is often strongly bound and thus
costs only a small amount of energy.

To consider configurations that violate $CP$, we must address the
question of whether to sum the quantity $-\half |\omega|$ over all
modes or the quantity $\omega$ the negative modes.  (If $C$ or $CP$ is
not broken, the two methods give the same answer.)  The former shows
more consistency with the boson computation, while the latter is what
one would expect from a Dirac sea picture, since the negative modes
are the ones that are filled.  The first method is correct.  One can
prove this formally \cite{Herbert}, but here we will show it via a
simple example, ordinary QED.  Consider QED with a massive fermion,
and imagine turning on a very weak, slowly varying electric field.
To leading order in the coupling, this field will shift both positive
and negative energy levels in the same direction, with corresponding particle
and antiparticle states being shifted by the same amount.  If we were
to sum only the shifts in the negative energy levels, we would find a
divergent result at the leading order in the Born approximation.
However, there is no available counterterm to cancel this infinity.
If instead we calculate the shift summing all of the levels, to first
order we simply get zero, since the shifts in the particle states
cancel against the shift in the antiparticle states.  This result
agrees with the loop calculation, where we find that the leading
contribution is identically zero, with no contribution from the
counterterm.  (Note that this result is different from what we find to
leading order for scalar field.  In that case, the particle and
antiparticle levels move toward zero, and both sums give the same
divergent result.  However, in this case there is a counterterm
available to cancel out this divergence.)

A hedgehog configuration breaks rotational invariance, but we can
choose it to stay within the spherical ansatz, so that it is invariant
under grand spin (simultaneous rotations in physical space and isospin
space).  We decompose the spectrum into a sum over channels labeled by their
grand spin and their parity.  Each channel now has two coupled degrees of
freedom instead of one, but still can be described by a second-order
(matrix) differential equation.  Summing the two eigenphases gives us
the total contribution to the density of states from that channel.
This matrix equation must also be iterated through fourth order to
obtain the Born approximations corresponding to the divergent diagrams.

Calculation in this generalized model in principle requires no further
modifications to the framework presented above.  However, there are
practical difficulties associated with the increased computational
complexity.  Even in the model without isospin, the third- and fourth-
order diagrams involve multidimensional integrals over external
momenta and Feynman parameters.  One possible simplification is to use
a ``toy'' boson model with the same divergence structure as the third-
and fourth-order diagrams.  If chosen properly, this model will
exhibit the same local at second order as the fermions have at third
and fourth order, so subtracting its second-order phase shifts and
adding back in its second-order diagram (in the fermionic
renormalization scheme, of course) can regulate the theory in a
computationally simpler way.  Such a manipulation is possible only
because the divergences are simple, local functions of the background
field; the finite parts of the bosonic diagram will of course be very
different from the finite parts of the fermionic diagrams.  However,
because we add back in everything we have subtracted, using
counterterms fixed in the fermion theory, we have not changed the
final answer.  Work is underway on this model \cite{future}.

\clearpage
\newpage

\chapter{Conclusions}

We have seen that the continuum density of states formalism has
allowed us to apply the powerful calculational tools of quantum
mechanics --- phase shifts, Born approximations and Levinson's theorem
--- to nontrivial problems in quantum field theory.  These formal tools
provide the robust framework we need to do unambiguous calculations.
Without them, we could have easily missed the subtleties of bosonic
and fermionic spectra that we have seen above, and such a mistake
would lead to a drastic change in the end result.  Such errors
are intolerable because of their theoretical consequences, as in the
case of the saturation of the BPS bound in 1+1 dimensional supersymmetric
theories, and because of their phenomenological consequences, as in
the case of heavy quark ``bags'' in the Standard Model.

One can envision many other applications of this work.  As mentioned
above, the most immediate is to include fermions coupled to scalars
with isospin in the spherical ansatz.  One could also imagine adding
gauge fields to such a model, to bring it still closer to the real
Standard Model.  One could then apply these methods to other gauge
theory solitons, such as 't Hooft-Polyakov monopoles or flux tubes in
superconductors.  It would also be interesting to see if one could use
these techniques to compute quantum corrections to the action of
instantons.

\clearpage
\newpage

\appendix
\chapter*{Appendix A:  Properties of reflectionless potentials in one
dimension}

In this section we review the properties of solutions of
\begin{equation}
\left(-\frac{d^2}{dx^2} + V_\ell(x)\right)\eta(x) = k^2 \eta(x)
\label{lHam}
\end{equation}
with
\begin{equation}
V_\ell(x) = -\ell(\ell+1)m^2 {\rm sech}^2 mx
\end{equation}
for an integer $\ell$.  For more details, see \cite{MF}.
The results used in the soliton problems above will then be obtained
by rescaling $x \to \frac{x}{\ell}$.
The basic technique will be to define raising and lowering operators
\begin{eqnarray}
a_\ell &=& i\left(\frac{d}{dx} + \ell m\tanh mx \right)\cr
a_\ell^\dagger &=& i\left(\frac{d}{dx} - \ell m\tanh mx\right)
\end{eqnarray}
so that we can rewrite eq.~(\ref{lHam}) as
\begin{equation}
\left(a_\ell^\dagger a_\ell - \ell^2 m^2\right)\eta(x) = k^2 \eta(x) =
\left(a_{\ell+1} a_{\ell+1}^\dagger - (\ell+1)^2 m^2\right)\eta(x).
\label{opHam}
\end{equation}
We can then connect the solutions for a given $\ell$ with those for
$\ell - 1$ by observing that eq.~(\ref{opHam}) implies that if $\tilde
\eta(x)$ is an eigenstate in the potential $V_{\ell-1}$, then $\eta(x)
= a_\ell^\dagger \tilde \eta(x)$ is an eigenstate in the potential
$V_\ell$ with the same eigenvalue $k^2$.  Thus we have
\begin{eqnarray}
\omega_\ell \tilde\eta(x) &=& a_\ell \eta(x) \cr
\omega_\ell \eta(x) &=& a_\ell^\dagger \tilde\eta(x)
\end{eqnarray}
with $\omega_\ell = \sqrt{k^2+l^2 m^2}$.  Thus the spectra of $V_\ell$ and
$V_{\ell-1}$ are identical, except that the spectrum of $V_\ell$ might contain
additional states annihilated by $a_\ell$.  Indeed, there is exactly
one such state, at $k^2 = -\ell^2 m^2$ (which becomes the zero mode of the
corresponding soliton).

By iterating this process, we arrive at $\ell=0$, which is a free
particle.  Since we know how to solve that problem completely, we can
go backwards and solve for any $V_\ell$ simply by applying raising
operators and at each step solving for the one additional zero mode.

For example, for $\ell = 1$ we start with the free scattering states
$e^{ikx}$ and apply $a^\dagger_1$, giving
\begin{equation}
\eta^1_k(x) = \frac{i}{\omega_1}\left(\frac{d}{dx} - m\tanh mx\right)
e^{ikx} =  -\frac{1}{\omega_1} (k + i m\tanh mx) e^{ikx}
\end{equation}
and a new bound state at $k^2 = -1$ that satisfies
\begin{equation}
i\left(\frac{d}{dx} + m\tanh mx\right) \eta^1_0(x) = 0 \rightarrow
\eta^1_0(k) = \frac{1}{\sqrt 2} {\rm sech} mx.
\end{equation}
We can then proceed to $\ell=2$, obtaining scattering states
\begin{eqnarray}
\eta^2_k(x) &=& \frac{i}{\omega_2}\left(\frac{d}{dx} - 2 m\tanh
mx\right) \eta^1_k(x) \cr
&=&  \frac{1}{\omega_1 \omega_2} ((k + 2im\tanh mx)(k + im \tanh mx)+ m^2 {\rm
sech}^2 mx)e^{ikx} \cr
&=& \frac{1}{\omega_1 \omega_2} (m^2 + k^2 + 3imk\tanh mx -
3m^2\tanh^2 mx)e^{ikx} 
\end{eqnarray}
a bound state at $k^2 = -1$ given by 
\begin{equation}
\eta^2_1(x)  = \frac{i}{\sqrt{3}}\left(\frac{d}{dx} - 2 m\tanh
mx\right) \eta^1_0(x)  = -i\sqrt{\frac{3}{2}} {\rm sech} mx \tanh mx 
\end{equation}
and a new bound state with $k^2 = -4$ solving
\begin{equation}
i\left(\frac{d}{dx} + 2m\tanh mx\right) \eta^2_0(x) = 0 \rightarrow
\eta^2_0(k) = \sqrt\frac{3}{4} {\rm sech}^2 mx.
\end{equation}

We can extend to any integer $\ell$ by repeating this process.
We find that the scattering from these potentials is reflectionless,
with phase shift
\begin{equation}
\delta_{\ell}(k) = 2\sum_{j=1}^{\ell} \tan^{-1}\left(\frac{jm\ell}{k} \right)
\label{phaseshift}                          
\end{equation}
and bound states at
\begin{equation}
k^2 = -(mj\ell)^2
\end{equation}
for $j=0,\dots,\ell$.  The state at $k^2=0$ is a ``half-bound'' state,
right at threshold, which goes to a constant as $x\to\pm\infty$.
This state is given by acting with our raising operators on the
threshold state in the free case, which is just a constant wavefunction.
Thus for $\ell = 1$ we find
\begin{equation}
\eta^1_{\rm thresh} = \frac{i}{m}\left(\frac{d}{dx} - m\tanh
mx\right) 1 = -i\tanh mx
\end{equation}
and for $\ell = 2$ we obtain
\begin{eqnarray}
\eta^2_{\rm thresh}(x) &=& \frac{i}{2m}\left(\frac{d}{dx} - 2 m\tanh
mx\right) \eta^1_{\rm  thresh} (x) \cr
&=& \half(1  - 3\tanh^2 mx).
\end{eqnarray}
We see that in both cases, these states are the $k\to 0$ limit of the
scattering states, and have the property that they go to a constant
(rather than a straight line of some nonzero slope) as $x\to\pm\infty$.

Since $V_\ell$ is symmetric, we can separate the spectrum of
wavefunctions into symmetric and antisymmetric channels.  The symmetry
of the bound states will alternate, with the lowest energy bound state being
symmetric.  The phase shift can also be decomposed into contributions
from the two channels, with
\begin{equation}
  \delta_{\ell}(k) = \delta^{\rm S}_{\ell}(k) + \delta^{\rm A}_{\ell}(k).
\end{equation}
That the scattering is reflectionless is equivalent to
\begin{equation}
\delta^{\rm S}_{\ell}(k) = \delta^{\rm A}_{\ell}(k).
\label{symantiequal}
\end{equation}
Levinson's theorem for the two channels gives \cite{lev}
\begin{eqnarray}
\delta^{\rm A}_{\ell}(0) &=& \pi n^A_\ell \cr
\delta^{\rm S}_{\ell}(0) &=& \pi (n^S_\ell - \half)
\label{symantiLev}
\end{eqnarray}
where $n^A_\ell$ and $n^S_\ell$ are the numbers of antisymmetric and symmetric
bound states in partial wave $\ell$, with threshold states counted as
one half.  Thus we see that the threshold states are essential in reconciling
eq.~(\ref{symantiequal}) with eq.~(\ref{symantiLev}).

\clearpage
\newpage

\chapter*{Appendix B: Phase shifts in arbitrary dimensions}

In this section we demonstrate explicitly the link between the leading
Born approximation to the phase shifts and tadpole diagrams by
computing both in dimensional regularization.  We consider only the
Bose case for simplicity.

In $n$ spatial dimensions, we consider the bosonic small oscillations
in partial wave $\ell$ with wave number $k$ in the presence of a
$n-$dimensionally spherically symmetric bosonic potential $V(x)$.  As
in three dimensions, we can separate variables and obtain a differential
for the wavefunction $\eta(r)$ in partial waves $\ell$.  Also as in
three dimensions, it is convenient to use a rescaled wavefunction $u =
r^{\frac{n}{2} - 1} \eta$, giving a generalized Bessel's equation
\begin{equation}
-\left(\frac{d^2}{dr^2} +\frac{1}{r} \frac{d}{dr} -
\frac{1}{r^2}\left(\ell(\ell + n - 2) + (1-\frac{n}{2})^2
\right)\right) u + Vu = k^2 u.
\label{appb1}
\end{equation}
This is the same equation as in three dimensions except
that the eigenvalue of the Casimir operator $L^2 = \half
M_{\alpha\beta} M^{\alpha\beta}$ has been generalized from $\ell (\ell+1)$ to
$\ell (\ell + n - 2)$.

The degeneracy factor for this partial wave (i.e. the dimension of
this representation of the rotation group, which gives the generalization
of $2\ell +1$ in three dimensions) is the dimension of the space of
symmetric tensors with $\ell$ indices that each run from 1 to $n$ with
all traces (contractions) removed.  Working out the combinatorics gives
\begin{equation}
N_\ell^n = \frac{(n+\ell-1)!}{\ell!(n-1)!} -
\frac{(n+\ell-3)!}{(\ell-2)!(n-1)!}
\label{appb2}
\end{equation}
which we will analytically continue to arbitrary $n$ using the gamma
function, which satisfies
\begin{equation}
x\Gamma\left(x\right) = \Gamma\left(x+1\right) 
\hbox{\qquad and \qquad} \Gamma\left(n+1\right) = n!
\end{equation}
For a fuller derivation of eq.~(\ref{appb1}) and eq.~(\ref{appb2}) see
for example Appendix B of \cite{Seiberg}.

As we have seen above, the tadpole graph requires the external
momentum to be equal to zero.  Thus we should expect that
both the leading Born approximation and the tadpole graph will
depend only on the spatial average of the potential (its $p=0$
component),
\begin{equation}
\langle V \rangle = \frac{2\pi^\frac{n}{2}}{\Gamma\left(\frac{n}{2}\right)}
\int_0^\infty V(r) r^{n-1}  \,dr.
\end{equation}
The leading Born approximation to the phase shift is
\begin{equation}
\delta_\ell^{(1)}(k) = - \frac{\pi}{2} 
\int _0^\infty J_{\frac{n}{2} +\ell-1}(kr)^2 V(r) r \, dr
\end{equation}
where $J_\nu(z)$ is an ordinary Bessel function, related to the
spherical Bessel function by $j_p(z) = \sqrt{\frac{\pi}{2z}} J_{p+\half}(z)$.
The leading correction to the energy is
\begin{equation}
{\cal E}^{(1)} = 
\sum_{\ell=0}^{\infty} N_\ell^n
\int_0^\infty \frac{dk}{2\pi} (\omega - m) \frac{d\delta_\ell^{(1)}(k)}{dk}
\end{equation}
where we have used Levinson's theorem as in eq.~(\ref{unrenormprime}).
This manipulation was not necessary in our calculations in three
dimensions since we did not have any singularities at $k=0$, but it
still would have been a perfectly valid thing to do; all the Born
approximations vanish at both $k=0$ and $k=\infty$.  However, to
calculate in arbitrary dimensions we will need to perform the $\ell$
sum first.  Thus, to avoid the same infrared singularities we saw in one
dimension, we again have used Levinson's theorem to add a total derivative
to the unregulated integral.  As a result, our analysis of arbitrary
dimensions will provide further evidence that the correct form of the
unregulated sum is indeed that of eq.~(\ref{unrenormprime}).

Using the Bessel function identity
\begin{equation}
\sum_{\ell=0}^{\infty}
\frac{(2q+2\ell)(2q+\ell-1)!}{\ell!}J_{q+\ell}(z)^2 = \frac{(2q)!}{(q!)^2}
\left(\frac{z}{2}\right)^{2q}
\end{equation}
we can explicitly do the sum over $\ell$, giving
\begin{equation}
{\cal E}^{(1)} = -\frac{\langle V \rangle }{(4\pi)^\frac{n}{2}
\Gamma\left(\frac{n}{2}\right)} \int_0^\infty (\omega - m)(n-2) k^{n-3} \, dk.
\end{equation}
The $k$ integral can be calculated near $n=\half$ and then
analytically continued, giving
\begin{equation}
\int_0^\infty (\omega - m) k^{n-3} \, dk = - \frac{m^{n-1}}{4 \sqrt{\pi}}
\Gamma\left(\frac{1-n}{2}\right) \Gamma\left(\frac{n-2}{2}\right)
\end{equation}
and so we find
\begin{equation}
{\cal E}^{(1)} = \frac{\langle V \rangle }{(4\pi)^\frac{n+1}{2}}
\Gamma\left(\frac{1-n}{2}\right) m^{n-1}
\end{equation}
in agreement with what we obtain using standard dimensional
regularization of the tadpole diagram in $n+1$ space-time dimensions.
\bibliography{main}
\bibliographystyle{plain}

\end{document}